# NEW RESULTS FOR RADIATIVE $^3$He($^2$H,γ)$^5$Li CAPTURE AT ASTROPHYSICAL ENERGY AND ITS POSSIBLE ROLE IN ACCUMULATION OF $^6$Li AT THE BBN

S. B. Dubovichenko,[1,2] N. A. Burkova,[2] A. V. Dzhazairov-Kakhramanov,[1] Tkachenko A.S.[1,2], Kezerashvili R.Ya.[3,4], Zazulin D.M.[2,5]

[1] Fesenkov Astrophysical Institute "NCSRT" ASA MDASI RK, 050020, Almaty, Kazakhstan
[2] al-Farabi Kazakh National University, 050040, Almaty, Kazakhstan
[3] Physics Department, New York City College of Technology, City University of New York, 300 Jay Street, Brooklyn, New York 11201, USA 5
[4] Graduate School and University Center, City University of New York, New York 10016, USA
[5] Institute of Nuclear Physics, ME RK, 050032, Almaty, Kazakhstan

## ABSTRACT

Big Bang Nucleosynthesis (BBN) relevance reactions $^3$He($^2$H,γ)$^5$Li, $^3$H($^3$He,γ)$^6$Li, $^5$Li($n$,γ)$^6$Li as a key to approach for scenario of $^6$Li formation are treated. The rates of reaction for these processes are analyzed. Comparison of the reactions rates and the prevalence of light elements leads to the assumption that the two-step process $^2$H + $^3$He → $^5$Li + γ and $n$ + $^5$Li → $^6$Li + γ can make a significant contribution to the formation of $^6$Li at the BBN at least at temperatures $T_9$ of the order of unity.

Calculations of the total cross sections, astrophysical $S$-factor, and reaction rates have been performed for $^3$He($^2$H,γ)$^5$Li radiative capture within the modified potential cluster model with forbidden states, which follow from the classification of the orbital cluster states according to Young diagrams. Numerical data and corresponding parametrizations cover the energy range up to 5 MeV and temperature range $T_9$<10. An updated compilation of detailed data for the reaction $^3$He($^2$H,γ)$^5$Li are presented.

## 1. INTRODUCTION

The interest to the radiative capture reactions in the isobar-analogue channels $^3$He($^2$H, γ)$^5$Li and $^3$H($^2$H, γ)$^5$He is primarily due to the following two reasons: these reactions are parts of nucleosynthesis chain of the processes occurring in the early stages of a stable star formation, as well as possible candidates for the overcoming of the well-known problem of the $A = 5$ gap in the synthesis of light elements in the primordial Universe (Barnes 1982) and application of these processes for the diagnostics of nuclear fusion efficiencies of $^2$H($^3$H, $n$)$^4$He and $^2$H($^3$He, $p$)$^4$He reactions used for experimental studies of tokamak plasmas (Lierrer et al. 1992; Sharapov et al. 2016).

It is believed that the $^3$He($^2$H, γ)$^5$Li process is well studied experimentally. Since 1954 (Blair et al. 1954) a sufficient number of experimental works, see, for example, Kraus et al. (1968), Del Bianco et al. (1968), Buss et al. (1968) Schroder and Mausberg (1970), Weller and Balbes (1998), are devoted to measuring these process, and the most recent data are presented by Balbes et al. (1994), including compilations of experimental data on cross sections, astrophysical factors and rates of this reaction at energies less than 200 keV (Sharapov et al. 2016; Kiptily et al. 2006). However, in our opinion, the experimental and theoretical situation is far from unambiguous and requires systematic analysis, which we implement in this article relying as a certain criterion on our model theoretical calculations.

There is another "unambiguous" opinion: due to the smallness of the cross section the $^3$He($^2$H,$\gamma$)$^5$Li reaction does not contribute to the astrophysical processes (Caughlan & Fowler 1988). In this article we address this issue and demonstrate that this statement is disputable, since the rate of this reaction is not negligible. In addition, we will consider a possible scenario for astrophysical processes of $^6$Li formation involving a short-lived $^5$Li isotope.

In this paper, the radiative $^3$He($^2$H,$\gamma$)$^5$Li capture is considered on the basis of the modified potential cluster model (MPCM) and new results are obtained for dipole $E$1 and $M$1 transitions, taking into account the mixing of the doublet and quartet spin channels, for both the scattering and the bound ground states. We construct the potentials of the intercluster interaction based on the description of the known scattering phase shifts and the main characteristics of the ground state (GS) of $^5$Li. The total cross sections for the $^3$He($^2$H,$\gamma$)$^5$Li capture rate into the GS of $^5$Li at energies in the frame of the center-of-mass (c.m.) from 5 keV to 5 MeV are calculated. Finally, we propose simple analytic parametrizations for the astrophysical $S$-factor and the rate of this reaction.

The parametrization of the cross sections for the $^3$H($^3$He,$\gamma$)$^6$Li and $^5$Li($n$,$\gamma$)$^6$Li processes of radiative capture is carried out. The rates corresponding to these processes were calculated, their parametrization was performed, and a comparison with the $^3$He($^2$H,$\gamma$)$^5$Li and $^4$He($^2$H,$\gamma$)$^6$Li capture reactions rate are made. On the basis of comparisons of the rates of these reactions and the prevalence of light elements, we suggests that the two-step process $^2$H + $^3$He → $^5$Li + $\gamma$ and $n$ + $^5$Li + $\gamma$ → $^6$Li + $\gamma$ can make a certain contribution to the production of $^6$Li at the BBN at least at temperatures $T_9$ of the order of unity. In this temperature range the number of neutrons does not yet begin to decrease, while the number of $^2$H and $^3$He nuclei is already reaching its maximum, which leads to increase in the reaction yield $^2$H + $^3$He → $^5$Li + $\gamma$.

This paper is organized in the following way. In Sec. 2 we review of experimental results and discuss the application range of the $^3$He($^2$H,$\gamma$)$^5$Li capture reaction. The model and methods of calculation and potentials for elastic $^2$H$^3$He scattering and $^2$H$^3$He bound states are presented in Sec. 3. In Sec. 4 results of calculations for total cross sections, astrophysical $S$-factor and reaction rate are presented and discussed. In Sec. 5 the two-step process $^2$H + $^3$He → $^5$Li + $\gamma$ and $n$ + $^5$Li +$\gamma$ → $^6$Li + $\gamma$ is considered and discussed as an alternative way of the formation of $^6$Li at the BBN. Conclusions follow in Sec. 6.

## 2. APPLICATION RANGE OF THE $^3$He($^2$H,$\gamma$)$^5$LI CAPTURE REACTION AND REVIEW OF EXPERIMENTAL RESULTS

Let us now examine in more detail the various aspects of the $^3$He($^2$H,$\gamma$)$^5$Li reaction, including the experimental data presented in the data base (EXFOR, 2013) and the original references cited therein, and the role of this process in various fields of application.

### 2.1. Nuclear astrophysics aspects

Our parametrization of the experimental data for $^2$H capture in $^3$He (Buss et al. 1968) for the $S$-factor at energies from 0.2 to 1.0 MeV according to Breit-Wigner and its further extrapolation to zero energy leads to a value of 0.24 keV·b. Let us give for comparison several known values of the *S*-factors at zero energy for some radiative capture reactions. For example, the latest data for the astrophysical factor of the proton capture on $^2$H give the value $S(0) = 0.216(11) \cdot 10^{-3}$ keV·b (Casella et al. 2002), other results lead to $0.166(14) \cdot 10^{-3}$ keV·b (Schimd et al. 1997). At the same time, for the proton capture in $^3$H, $2.0(2) \cdot 10^{-3}$ keV·b (Canon, R., et al. 2002) is known, *i.e.* an order of magnitude greater. Cecil, et al. (1993), reported 1.3(3) keV·b for $^3$H($^2$H,γ)$^5$He capture, and for $^2$H($^2$H,γ)$^4$He capture a small value of $6.0(1.2) \cdot 10^{-6}$ keV·b. It is due to the strong *E*1 transition in this process which is forbidden due to equal masses of the particles in the initial channel (Sharapov et al. 2016).

Cecil et al. (1993), reported 0.36(9) keV·b for the astrophysical factor of the $^3$He($^2$H,γ)$^5$Li capture. At the same time, more recent results (Kiptily et al. 2006) for the $^3$He($^2$H,γ)$^5$Li capture for the *S*-factor is given the value 0.26(7) keV·b, which is in a good agreement with our value of 0.24 keV·b, which we report below. Although the error bands of these data (Kiptily, et al. 2006) and (Cecil et al. 1993) overlap, in fact, the values of the astrophysical factor of this reaction can be in the range 0.19–0.45 keV·b. This is a very large uncertainty for its values, which leads, despite the relatively low abundance of $^3$He, to the rather large uncertainty of the contribution of the reaction under consideration to the primordial nucleosynthesis of the Universe and the processes of initial star formation.

Therefore, a more detailed study of this process will allow to obtain the results for the reaction rate and to compare it with the rates of other reactions of thermonuclear fusion, given, for example, by Caughlan and Fowler (1988). This work contains simple parameterizations of rates for dozens of thermonuclear reactions, but there are no data for $^3$He($^2$H, γ)$^5$Li capture. Perhaps this is due to the fact that the rate of the reaction under study was considered negligible. However, below is shown that the rate of this reaction at certain temperatures is more than the rate of proton capture on $^2$H and is comparable with the rate of proton capture on $^3$H. Thus, the $^3$He($^2$H, γ)$^5$Li reaction rate among the rates of the processes of radiative capture in the considered temperature range of 0.01–10.0 $T_9$ is actually of the similar magnitude to the other processes which are considered as an important.

## *2.2 Nuclear Physics aspects and experimental results*

From the point of view of nuclear physics, practically all experimental studies of the total cross sections of $^3$He($^2$H, γ)$^5$Li reaction for capturing to the ground state of $^5$Li from the 1960s to the present are limited to two works published in late 1960s and early 1970s (Buss et al. 1968; Schroder & Mausberg 1970). It is only in these works the total cross sections of the radiative $^3$He($^2$H, γ)$^5$Li capture at the astrophysical energy range are shown in an explicit form.

In the first one, measurements were made at a deuteron energy of 200–1360 keV in the laboratory system (l.s.). For the $^3$He($^2$H,γ)$^5$Li reaction separation of the transitions into the GS and the first excited state (FES) of $^5$Li (see Figure 1) have been studied (Buss et al. 1968). However, as a result of the poor quality of the experimental data for the transition to the FES of $^5$Li, the integral cross sections are presented only for the transition to the ground state with a value of 21(4) μb for $\sigma_{\gamma 0,cm}$ at $E_R = E_{cm} = 0.27$ MeV (Buss et al. 1968). The angular distributions measured by Buss et al. (1968) at $E_{cm} = 0.288$ and 0.615 MeV for the transition to the ground state of $^5$Li was consistent with isotropy to within 10%. The total cross-sections obtained by Buss et al. (1968) with an accuracy of 20% are shown below in Figure 2 by green squares.

The total cross sections were measured by Schroder & Mausberg (1970) at excitation energy $E_x$ from 17.4 to 21.1 MeV. If we use the binding energy for the $^2$H$^3$He channel of 16.66

MeV (Tilley et al. 2002), we obtain capture energy of 0.74–4.44 MeV in the cm. Schroder and Mausberg (1970) used the value of the binding energy of 16.4 MeV, which is close to 16.39 MeV given by Ajzenberg-Selove (1988). Using energy of 16.4 MeV, we obtain capture energy of 1.0–4.7 MeV in the cm. We use these values throughout our work.

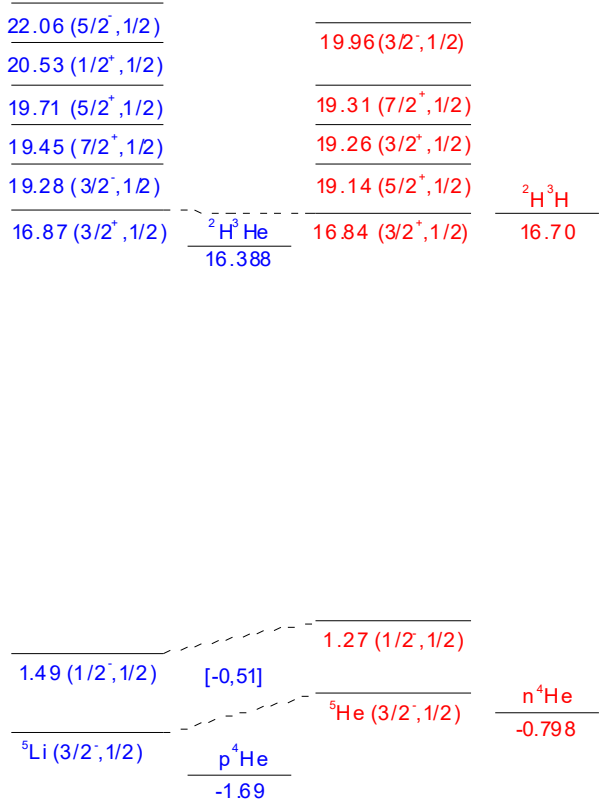

**Figure 1.** Experimental energy levels for $^5$Li and $^5$He from (Tilley et al. 2002), but with a binding energy of $^5$Li in the $^3$He$^2$H channel of 16.39 MeV and a binding energy of $^5$He in the $^3$H$^2$H channel of 16.70 MeV (Ajzenberg-Selove 1988).

The capture total cross sections obtained by Schroder and Mausberg (1970) have a wide maximum (the width of 3–4 MeV) with the value of 59(3.0) μb at the energy $E_x = 19.7$ MeV, i.e. at $E_{cm} = 3.3$ MeV. This maximum points to the existing group of broad levels in this energy range. This group of levels at energies 19.28–22.06 MeV is clearly visible in the spectra of $^5$Li (Tilley et al. 2002), shown in Figure 1. The total cross sections for capture into the GS obtained by Schroder and Mausberg (1970) are shown in Figure 2 by stars. The angular distributions at angles $\theta_{\gamma,\text{lab}} = 0°–130°$ obtained for the transition to the GS of $^5$Li were close to isotropic up to the energy of 4 MeV in cm. The statistical error of the data (Schroder & Mausberg 1970) was 5%, but due to the high uncertainty of the procedure of separating the peaks from the transitions to the ground and first excited states of $^5$Li the error in the determination of the absolute values of the cross sections can reach up to 40%.

Once again we note that the most complete databases of nuclear data such as (EXFOR, 2013), as well as well-known atomic characteristics databases, for example, PHYSICS, CDFE, NASA DATA (Fundamental Physical Constants, 2010; Nuclear Wallet Cards database, 2014), contain only these data for total cross sections of $^3$He($^2$H,γ)$^5$Li capture at low energies. Besides them, however, there are several publications (Blair et al. 1954; Kraus et al. 1968; Del Bianco et al. 1968; King et al. 1972; Balbes et al. 1994; Weller & Balbes 1998) in which measurements of the number of events observed in the experiment, polarizations, and differential capture cross sections are realized. Therefore, we perform a recalculation of some of the experimental measurements of these studies with the extraction of the total cross sections from them. Let us now explore in more detail on the experimental aspects of above mentioned studies and make a comparison of the total cross sections obtained with their help.

Probably, for the first time, the yields of the $^3$He($^2$H,γ)$^5$Li reaction at low energies were measured by Blair et al. (1954). It was assumed that the reaction proceeds with capture to the ground and first excited states of the nucleus. However, due to the insufficient energy resolution of the spectrometer, the peaks from these transitions could not be separated and as a result the total yield from all possible transitions was presented. The excitation function obtained for $\theta_{\gamma,\text{lab}} = 90°$ and $E_{\text{cm}} = 0.1–1.5$ MeV showed a wide resonance at $E_{\text{cm}} = 0.27$ MeV with a total cross section in the peak equal to 50(10) μb. Also Blair et al. (1954) measured the angular distribution of the total yield of the reaction at $E_{\text{cm}} = 0.35$ MeV, which turned out to be isotropic with an accuracy of 10%. To obtain the integral cross sections from the data (Blair et al. 1954) the yield of this reaction, shown in Figure 4 in this work was normalized to the total cross section measured at $E_R = E_{cm} = 0.27$ MeV. In Figure 2, the solid circles show the integral cross sections, which we recalculated from the yield of the $^3$He($^2$H,γ)$^5$Li reaction given by Blair et al. (1954)

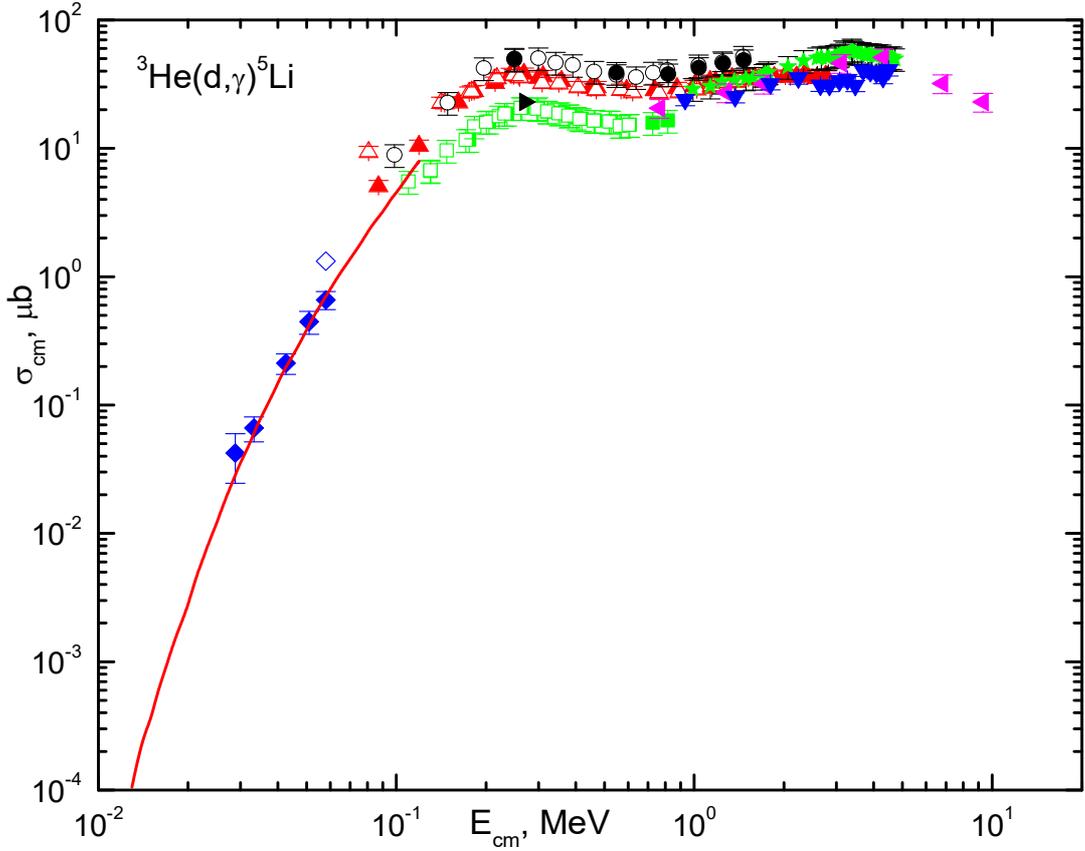

**Figure 2.** The total cross sections of the $^3$He($^2$H,γ)$^5$Li capture from (Buss et al. 1968; Schroder & Mausberg 1970) and recounted by us from some other *Ref*s, namely: ○ and ● – from (Blair et al. 1954) for $\gamma_0+\gamma_1$, □ and ■ – from (Buss et al. 1968) for $\gamma_0$, Δ and ▲ – from (Kraus et al. 1968) for $\gamma_0$, ▼ – from Ref. (Del Bianco et al. 1968) for $\gamma_0+\gamma_1$, ★ – from (Schroder & Mausberg 1970) for $\gamma_0$, ◄ – from Ref. (King et al. 1972) for $\gamma_0$, ♦ for $\gamma_0$ and ◊ for $\gamma_1$ – from (Cecil et al. 1985; Aliotta et al. 2001), ► – from (Weller & Balbes 1998; Balbes et al. 1994) for $\gamma_0$, the solid curve from (Kiptily et al. 2006) for $\gamma_0$.

Subsequently, Kraus et al. (1968) measured the differential cross sections of the $^3$He($^2$H,γ)$^5$Li reaction for $\theta_{\gamma,\text{lab}} = 90°$ in the energy range $E_{\text{cm}} = 0.08–2.76$ MeV. The authors of could not find any indications of the presence of a peak from the transition to the first excited state of $^5$Li in the spectra of the $^3$He($^2$H,γ)$^5$Li reaction, and the single broad peak in the spectra was interpreted as the peak from the transition to the ground state of $^5$Li. However, it is possible that this broad peak contained also the transition to the first excited state of $^5$Li but the accuracy of the experiment did not allow one to distinguish it. The measurement of the angular distributions at $E_{cm} = 0.3$ and 1.66 MeV with 7% error reported by Kraus et al. (1968) also turned out to be isotropic. The integral cross section obtained for $^3\text{He}(^2\text{H},\gamma)^5\text{Li}$ at the resonance energy $E_R = E_{cm} = 0.27$ MeV is 38(4) µb (Kraus et al. 1968).

Furthermore, the sum of the differential cross sections of the reaction $^2\text{H}(^3\text{He},\gamma)^5\text{Li}$ for the transitions to the ground and first excited states of $^5\text{Li}$ at $\theta_{\gamma,\text{lab.}} = 90°$ in the energy range of $E_{cm} = 0.93$–4.5 MeV were measured (Del Bianco et al. 1968). The excitation function obtained in this work, accurate to within 10% of the measurement by Del Bianco et al. (1968) demonstrates a smooth growth with increasing energy without a clear indication of the presence of any resonances of the reaction $^2\text{H}(^3\text{He},\gamma)^5\text{Li}$ in the considered energy range. When we obtained the integral cross sections from the data reported by Kraus et al. (1968) and Del Bianco et al. (1968), the differential cross sections shown in Figure 5 of (Kraus et al. 1968) and Figure 2 of (Del Bianco et al. 1968) were simply multiplied by $4\pi$ (here we assume the isotropy in the angular distributions that was shown in all works listed above). These results are shown in Figure 2 by triangles. It can be seen from Figure 2 that the data obtained by Kraus et al. (1968) differ quite strongly from the measurements done by Buss et al. (1968), although in both cases it is said that capture is only to the GS of $^5\text{Li}$. In addition, the results of Del Bianco et al. (1968) for capture to the GS and FES lie below the cross sections obtained by Schroder and Mausberg (1970) for the capture to the GS.

Somewhat later King et al. (1972) obtained the excitation functions of the reaction $^2\text{H}(^3\text{He},\gamma)^5\text{Li}$ at $\theta_{\gamma,\text{lab}} = 90°$ and $E_{cm} = 0.76$–10.1 MeV for transitions to the ground and first excited states of $^5\text{Li}$. Moreover in this work the sum of the differential cross sections for the transitions to the ground and first excited states of $^5\text{Li}$ at $\theta_{\gamma,\text{lab.}} = 90°$ is given. For five energies, the angular distributions for these transitions were measured. King at al. (1972) presented a table with the ratios of the coefficients of the Legendre functions of $A_2/A_0$ and $A_1/A_0$ for the cases when the first two and first three terms of the expansion in describing these angular distributions are taking into account. The error in determining the absolute values of the cross sections without taking into account the error in the procedure of separating of the peaks from the two transitions was 17%. A change of the angular distributions with the energy and the form of the excitation functions of the $^2\text{H}(^3\text{He},\gamma)^5\text{Li}$ reaction allowed the authors (King, et al. 1972) to make a conclusion about the existence of a broad resonance structure with the width of 3–4 MeV in the range of $E_{cm} = 0.76$–10.1 MeV and the maximum $\sigma_{\gamma 0} = 51(8)$ µb located at $E_{cm} \approx 4.3$ MeV. Using these data (King et al. 1972), we obtained the values of the integral cross section for $E_{cm} = 0.764, 1.282, 1.712, 3.112, 4.312, 6.812, 9.312$ MeV only for capture to the ground state of $^5\text{Li}$. since the These data (King et al. 1972) are unreadable for the first excited state of $^5\text{Li}$ due to the error of the order of 100%. We obtain the cross section for the energy $E_{cm} = 0.764, 1.282,$ and 1.712 MeV, by simply multiplying the differential cross sections (King et al. 1972) taken from Figure 8 of this work by $4\pi$, since at these energies the angular distributions are isotropic. We obtain the integral cross sections at $E_{cm} = 3.112, 4.312, 6.812, 9.312$ MeV, using data presented in Figure 8 and Table 1 (King et al. 1972). As a first step we find the coefficients $A_0$, and then these coefficients are multiplied by $4\pi$. The calculated integral cross sections for the two cases from Table 1 (King, et al. 1972) agree within their respective margins of error. These results are shown in Figure 2, which lie somewhat lower than the data given by Schroder and Mausberg (1970) for capture to the GS, but higher than the results reported by Del Bianco et al. (1968), where the sum of the cross sections for transitions to the GS and FES of $^5\text{Li}$ were measured.

Using the polarized deuteron beam with $E_{cm} = 0.48$ MeV, a thick target of $^3$He, which completely absorbed the beam energy, the angular distribution of the products of the reaction of deuteron capture by the $^3$He to the ground state of $^5$Li was measured (Balbes et al. 1994). The obtained angular distribution of γ quanta was isotropic within the error of 10%. In a similar work Weller and Balbes (1998) are presented data for a polarized deuteron beam with $E_{lab} = 0.6$ MeV and a $^3$He target in which the deuterons lost 0.3 MeV. Within the experimental error, the differential cross sections reported by Balbes et al. (1994) and Weller and Balbes (1998) coincide. The total cross section, obtained from these differential cross sections is 23 μb and depicted in Figure 2 in position of the first resonance of the reaction, *i.e.* at $E_R = E_{cm} = 0.27$ MeV.

To conclude this review we note that the aforementioned difference in the energy of the channel reported by Tilley et al. (2002) and Ajzenberg-Selove (1988) force us to recalculate this energy. We used data for the masses of $^2$H and $^3$He 1875.613 MeV and 2808.392 MeV, respectively, from (Fundamental Physical Constants, 2010), and the mass of $^5$Li (4667.617 MeV) was taken from the database of (Nuclear Wallet Cards database, 2014). Then, for the binding energy of the $^3$He$^2$H channel of $^5$Li, a value of 16.388 MeV is obtained, which, within precision of rounding errors, coincides with 16.39 MeV given by Ajzenberg-Selove (1988). As we have already mentioned, a value of 16.66 MeV presented by Tilley et al. (2002) differs significantly from our results and the data of the survey given by Ajzenberg-Selove (1988). For $^5$He in the $^3$H$^2$H channel with $m(^3$H$) = 2808.921$ MeV (Fundamental Physical Constants, 2010) and $m(^5$He$) = 4667.838$ MeV (Nuclear Wallet Cards database, 2014), we have obtained $E_{bin.} = 16.696$ MeV, which is in good agreement with 16.70 MeV reported by Ajzenberg-Selove (1988). At the same time, the value of 16.792 MeV given by Tilley et al. (2002) also differs noticeably from our results and the data of Ajzenberg-Selove (1988).

### *2.3 Application to plasma*

Now let us consider another experimental studies related to the plasma of synthesis reactors performed by Cecil et al. (1993), Cecil et al. (1985a) and Cecil et al. (1985b), which are applicable when considering the plasma of synthesis reactors. Cecil et al (1993) presented an astrophysical *S*-factor at zero energy and the total cross section of the $^2$H($^3$He,γ)$^5$Li reaction for the capture to the GS at 40 keV in cm. In the other work (Cecil, 1985), the reaction rate of the $^2$H($^3$He,γ)$^5$Li capture to the GS is given and we use the latter rate further to compare with our results. The results for the total cross sections and the astrophysical *S*-factor for the capture to the GS at the energy range 13–120 keV in cm are given and the reaction rate at temperatures up to 1.0 $T_9$ is presented (Kiptily et al. 2006). In Figure 2, the integral cross sections (Kiptily et al. 2006) for nuclei not screened by the electron shells of atoms are depicted by the solid curve. These results (Kiptily et al. 2006) are also obtained for their application and use in the study of thermonuclear processes in the plasma of artificial thermonuclear fusion reactors.

All these results are based on the experimental work of (Cecil et al. 1985b), in which the $^3$He($^2$H,γ)$^5$Li reaction was studied at $E_{cm} = 0.025$–0.06 MeV. In this work, the branching ratios for the $^3$He($^2$H,γ$_0$)$^5$Li/$^3$He($^2$H,*p*)$^4$He at five energies within the aforementioned range and for the $^3$He($^2$H,γ$_1$)$^5$Li/ $^3$He($^2$H,*p*)$^4$He at $E_{cm} \approx 0.06$ MeV were measured within the errors range 12% to 40%. It is shown by Cecil, Cole & Philbin (1985) that the ratio of the branching of $^3$He($^2$H,γ$_0$)$^5$Li/ $^3$He($^2$H,*p*)$^4$He in the range of $E_{cm} = 0.025$–0.06 MeV is practically constant and equal to $(4.5\pm1.2)\cdot10^{-5}$, and for the case of $^3$He($^2$H,γ$_1$)$^5$Li/ $^3$He($^2$H,*p*)$^4$He at $E_{cm} \approx 0.06$ MeV the value of $(8 \pm 3)\cdot10^{-5}$ is obtained. Moreover, the authors (Cecil et al. 1985b) were carried out a simple extrapolation of the branching ratio for the $^3$He($^2$H,$\gamma_0$)$^5$Li/ $^3$He($^2$H,$p$)$^4$He to the point of 0 keV and the value of $(8 \pm 2) \cdot 10^{-5}$ was obtained. However, it is possible that the increased value of the branching ratio for the transition to the ground state obtained at the lowest energy (Cecil et al. 1985b) can be explained by the large error of about 40%, and that in fact the ratio of the branching for the $^3$He($^2$H,$\gamma_0$)$^5$Li/$^3$He($^2$H,$p$)$^4$He is constant up to 0 keV and equal to $(4.5 \pm 1.2) \cdot 10^{-5}$.

To obtain the integral cross sections of the reaction $^3$He($^2$H,$\gamma_{0,1}$)$^5$Li in the region $E_{cm}$ = 0.025–0.06 MeV, presented in Figure 2, we use data (Cecil et al. 1985b) and (Aliotta et al. 2001) given in Figure 4 and Table 1 respectively. Aliotta et al. (2001) presented the astrophysical $S$-factor of the $^3$He($^2$H,$p$)$^4$He reaction for $E_{cm}$ = 0.008–0.06 MeV with an error of about 3%. First, from the $S$-factors given by Aliotta et al. (2001) we obtain the integral cross sections for the energies considered by Cecil, Cole & Philbin (1985b), and then these cross sections we multiply by the values of the branching ratios for the $^3$He($^2$H,$\gamma_{0,1}$)$^5$Li/ $^3$He($^2$H,$p$)$^4$He given by Cecil et al. (1985b). Moreover, it can be seen from Figure 4 of Aliotta et al. (2001) work that in the range of $E_{cm}$ = 0.025–0.06 MeV, the $S$-factors of the $^3$He($^2$H,$p$)$^4$He reaction for screened by the electron shells of the target atoms nuclei differ from $S$-factors for bare nuclei by no more than 5%. Therefore, in constructing the experimental bare integrated cross sections for the $^3$He(d,$\gamma_{0,1}$)$^5$Li reactions with an accuracy of 13–41%, we do not take into account the screening effects.

## *2.4 Some theoretical results*

There are theoretical studies (Chwieroth et al. 1973; Kurath 1993; Tanifuji & Kameyama 1996; Wagner & Werntz 1971), which provide based on various approaches descriptions of either the differential cross sections and the polarization characteristics, or the spectral levels and their widths. Modern "*ab initio*" microscopic calculations for systems with $A$ = 5, as well as the reactions $^3$H($^2$H,$n$)$^4$He and $^3$He($^2$H,$p$)$^4$H in the context of applications to thermonuclear processes in the Big Bang and laboratory fusion are presented by Navratil & Quaglioni (2012). However, the processes of radiation capture, for example, $^3$He($^2$H,$\gamma$)$^5$Li the authors did not consider. Theoretical calculations that would include the main static and dynamic characteristics of the reaction $^3$He($^2$H,$\gamma$)$^5$Li and the final nucleus $^5$Li could not be found, at least in the sources available to us. As a result one can conclude that consideration of all the experimental and theoretical works presented above, the capture reaction $^3$He($^2$H,$\gamma$)$^5$Li does not seem to be sufficiently studied both in the experimental and theoretical senses.

Below we use our expertise and a lot of experience in using the single-channel modified potential cluster model (MPCM) with forbidden states (FS) and the classification of cluster states according to Young diagrams (Dubovichenko 2015a, 2016; Dubovichenko & Dzhazairov-Kakhramanov 2012, 2015, 2016, 2017; Dubovichenko, Dzhazairov-Kakhramanov & Afanasyeva 2017; Dubovichenko & Burkova 2014; Dubovichenko et al. 2014), and apply to study the radiative $^3$He($^2$H,$\gamma$)$^5$Li capture. This model is much simpler than the known Resonating Group Method (RGM) (Mertelmeir & Hofmann 1986) and its modifications (Descouvemont et al. 2013, 2014), but in many cases it allows one to obtain quite reliable numerical results for many reactions such as radiative capture. In particular, on the basis of a unified concept and calculation methods, MPCM allowed one to describe the main characteristics of already 30 thermonuclear processes of radiative capture at astrophysical and thermal energies (Dubovichenko 2015a, 2016).

# 3. MODEL AND SOME THEORETICAL METHODS

The modified potential cluster model, which is used in the present calculations, is described in detail in (Dubovichenko 2015a, 2016) and, in part, in (Dubovichenko & Dzhazairov-Kakhramanov 2012, 2015, 2016, 2017; Dubovichenko, Dzhazairov-Kakhramanov & Afanasyeva 2017; Dubovichenko & Burkova 2014; Dubovichenko et al. 2014). One of its modifications relies on the explicit dependence of the interaction potentials on Young orbital diagrams and on taking into account the mixing of cluster states according to these diagrams. The explicit dependence of the potentials on Young diagrams was taken into account already by Neudatchin et al. (1992). As a result, the wave functions (WF) of the relative motion of clusters may be mixed by orbital diagrams (Dubovichenko 2016; Dubovichenko & Dzhazairov-Kakhramanov 2015; Dubovichenko, Dzhazairov-Kakhramanov & Afanasyeva 2017; Neudatchin et al. 1992; Neudatchin, Sakharuk & Dubovichenko 1995). A similar situation was observed earlier, for example, in the $N^2H$ (Dubovichenko & Dzhazairov-Kakhramanov 2015; Neudatchin et al. 1992), $p^3H$ (Dubovichenko, Dzhazairov-Kakhramanov & Afanasyeva 2017; Neudatchin, Sakharuk & Dubovichenko 1995) or $^2H^3He(^2H^3H)$ systems (Neudatchin et al. 1992), affecting the interaction potentials, which are used in calculations, for example, of the radiative capture total cross sections.

## *3.1 General definitions*

All the general expressions for the calculation of various nuclear characteristics which we are using in the present work are given in our early papers (Dubovichenko 2015a, 2016). In particular, the expressions

$$\sigma_c(NJ,J_f) = \frac{8\pi Ke^2}{\hbar^2 q^3} \frac{\mu}{(2S_1+1)(2S_2+1)} \frac{J+1}{J[(2J+1)!!]^2} A_J^2(NJ,K) \cdot \sum_{L_i,J_i} P_J^2(NJ,J_f,J_i) I_J^2(J_f,J_i), \quad (1)$$

are used for the total cross sections (see, for example, (Dubovichenko & Dzhazairov-Kakhramanov 1997) or (Angulo et al. 1999)). In Eq. (1) $\mu$ is reduced mass in the initial channel, $q$ is the wave number in fm$^{-1}$ related to the c.m. energy as $q^2 = (2\mu E)/\hbar^2$, $S_1$, $S_2$ particle spins in the initial channel, $K$, $J$ are the wave number and the momentum of emitted $\gamma$ quantum, and $N$ denotes electric $E$ or magnetic $M$ transitions of rank $J$ from the initial state $J_i$ to the final one $J_f$.

For electric convection $EJ(L)$ transitions ($S_i = S_f = S$) there are the following expressions for $P_J$, $A_J$ and $I_J$ in (1) (Dubovichenko 2015a, 2016):

$$P_J^2(EJ,J_f,J_i) = \delta_{S_i S_f}[(2J+1)(2L_i+1)(2J_i+1)(2J_f+1)](L_i 0 J 0 | L_f 0)^2 \begin{Bmatrix} L_i & S & J_i \\ J_f & J & L_f \end{Bmatrix}^2,$$

$$A_J(EJ,K) = K^J \mu^J \left( \frac{Z_1}{m_1^J} + (-1)^J \frac{Z_2}{m_2^J} \right), \qquad I_J(J_f,J_i) = \left\langle \chi_f \left| R^J \right| \chi_i \right\rangle. \quad (2)$$

Here $S_i$, $S_f$, $L_f$, $L_i$, $J_f$, $J_i$ are the corresponding momenta in the initial and final states, $m_1$, $m_2$, $Z_1$, $Z_2$ are masses and charges in the initial channel, and $I_J$ is the overlapping integral over the radial functions for the scattering $\chi_i$ and bound $\chi_f$ states, depending on the cluster-cluster relative coordinate $R$.

For the spin dipole magnetic $M1(S)$ transition, i.e. for $J = 1$, the following expressions were obtained ($S_i = S_f = S$, $L_i = L_f = L$) (Dubovichenko 2015a, 2016)

$$P_1^2(M1, J_f, J_i) = \delta_{S_i S_f} \delta_{L_i L_f} \left[ S(S+1)(2S+1)(2J_i+1)(2J_f+1) \right] \begin{Bmatrix} S & L & J_i \\ J_f & 1 & S \end{Bmatrix}^2,$$

$$A_1(M1, K) = i \frac{\hbar K}{m_0 c} \sqrt{3} \left( \mu_1 \frac{m_2}{m} - \mu_2 \frac{m_1}{m} \right), \qquad I_J(J_f, J_i) = \langle \chi_f | R^{J-1} | \chi_i \rangle. \tag{3}$$

In Eq. (3) $m = m_1 + m_2$ is the mass of nucleus, $\mu_1$, $\mu_2$ are the magnetic moments of the clusters (Chart nucl. shape & size param. ($0 \leq Z \leq 14$), 2015), all other notations are the same as in (2). For the light clusters $\mu(^2H) = 0.857438$, $\mu(^3H) = 2.978662$, $\mu(^3He) = -2.127625$. The exact values for the particle masses use: $m(^2H) = 2.013553$ amu, $m(^3H) = 3.015501$ amu, and $m(^3He) = 3.014932$ amu (Fundamental Physical Constants, 2010). The constant $\hbar^2/m_0$ is equal to 41.4686 MeV fm$^2$, where $m_0$ is the atomic mass unit (amu). The Coulomb parameter $\eta = (\mu Z_1 Z_2 e^2)/(\hbar^2 q) = 3.44476 \cdot 10^{-2} (\mu Z_1 Z_2)/q$. The Coulomb potential for the point-like particles is of the form $V_c(\text{MeV}) = 1.439975(Z_1 Z_2)/R$. Here we are giving the numerical values as the accuracy of the further calculations depends strongly over them, especially in the resonance energy ranges, as well as the $^5$Li binding energy in $^2$H$^3$He cluster channel. Principles for constructing intercluster potentials, numerical calculation methods and some computer programs on FORTRAN-90 are given by Dubovichenko (2015a, 2016) and partly by Dubovichenko and Dzhazairov-Kakhramanov (2015, 2016).

### 3.2 Potentials for elastic $^2H^3He$ scattering

We present the classification by orbital symmetries of the $^2$H$^3$He and $^2$H$^3$H systems, i.e. a configuration of 2+3 nucleons. The doublet channel spin ($S = 1/2$) scattering states depend on the two allowed orbital Young diagrams {41} and {32}, and these are regarded as mixed in terms of the orbital symmetries. The quartet channel spin ($S = 3/2$) allows only one symmetry {32}, so these states are pure according to the Young diagrams. Therefore, it is assumed that states with a minimal spin in some lightest cluster systems scattering processes, including those considered here, can turn out to be mixed in Young orbital diagrams, as was shown by Neudatchin et al. (1992) and some previous works of these authors (see references in (Neudatchin et al. 1992)).

The classification of states according to Young diagrams for these systems is given in Table 1 and was obtained by Neudatchin et al. (1992) and Dubovichenko and Dzhazairov-Kakhramanov (1997) on the basis of general tables of Young diagrams products given by Itzykson and Nauenberg (1966). Table 1 shows the forbidden state with the Young diagram {5} for the $S$-waves of doublet and quartet channels, while in the $P$-wave the forbidden state with the Young diagram {41} is present only for the quartet channel, in the doublet channel this state is allowed.

At the same time, the states of clusters in the discrete spectrum, for example, the ground states of $^5$He and $^5$Li nuclei are assumed to be pure with the Young diagram {41} (Neudatchin et al. 1992). Furthermore, it is assumed that since the scattering states and the discrete spectrum depend on different Young diagrams, it is possible to juxtapose them different interaction potentials. In other words, explicit dependence of the potentials on the orbital symmetries {$f$} is allowed, and not only on the quantum numbers $JLS$ – total angular momentum, orbital angular momentum and spin of the nuclear particles system (Neudatchin et al. 1992). In more detail, the case with the mixing of cluster states by Young diagrams was described by us in recent publications (Dubovichenko & Dzhazairov-Kakhramanov 2015; Dubovichenko, Dzhazairov-Kakhramanov & Afanasyeva 2017).

Table 1. Classifications of the allowed (AS) and forbidden states (FS) in cluster systems with $A = 5$. Here $T$, $S$ and $L$ are isospin, spin and the orbital angular momentum of the particle system, $\{f\}_S$, $\{f\}_T$, $\{f\}_{ST}$ and $\{f\}_L$ are spin, isospin, spin-isospin, and possible orbital Young diagrams, respectively and $\{f\}_{AS}$, $\{f\}_{FS}$ are Young diagrams of allowed and forbidden orbital states.

| System | $T$ | $S$ | $\{f\}_S$ | $\{f\}_T$ | $\{f\}_{ST}$ | $\{f\}_L$ | $L$ | $\{f\}_{AS}$ | $\{f\}_{FS}$ |
|---|---|---|---|---|---|---|---|---|---|
| $^2$H$^3$He | 1/2 | 1/2 | {32} | {32} | {5}+{41}+{32}+{311}+ +{221}+{2111} | {5} {41} {32} | 0 1 0,2 | – {41} {32} | {5} – – |
| $^2$H$^3$H | | 3/2 | {41} | {32} | {41}+{32}+{311}+{221} | {5} {41} {32} | 0 1 0,2 | – – {32} | {5} {41} – |

We note that the total angular momentum $J = 3/2^-$ for the GS or $J = 1/2^-$ for the FES in $P$ waves of $^5$He or $^5$Li nuclei can also be obtained in the doublet spin channel of $^2$H$^3$He ($^2$H$^3$H) clusters with total spin $S = 1/2$ and in the quartet channel with $S = 3/2$. Therefore, the GS and FES of these systems are actually $^{2+4}P$ mixtures of such channels. These spin-mixed $^{2+4}P$ states turn out to be mixed also according to Young diagrams, since the doublet pure channel corresponds to the {41} diagram, and the quartet channel to the {32} one. However, now only one Young diagram corresponds to each spin channel, but not two, as it was in the previous case with mixing one spin channel in terms of orbital symmetries. Therefore, the potentials obtained below with $J = 3/2^-$ and $1/2^-$ for GS and FES are called pure, although in reality they are mixed along the spin, and, hence, along the Young diagrams.

In the present calculations of the nuclear characteristics of the reaction under consideration, the interaction potentials of particles have the form of a Gaussian attraction that widely used in our previous works

$$V(r, JLS\{f\}) = V_{0,JLS\{f\}} \exp(-\alpha_{JLS\{f\}} r^2) + V_c(r) \qquad (4)$$

with the Coulomb potential $V_c(r)$ of two point particles for the zero Coulomb radius that has the form which was given above.

Follow (Jenny et al. 1980) we considered the $^2$H$^3$He scattering phase shifts in the energy range up to 3 MeV in the c.m., which are sufficient for solving various astrophysical problems, in particular, for calculating the total cross sections for the capture reaction. Parameters of partial, i.e. depending on $^{2S+1}L_J$, potentials of the continuous spectrum, given in Table 2, including the resonance states potentials, which will be described in detail later. The characteristics of these states are given in the last column of Table 2 (Tilley et al. 2002; Ajzenberg-Selove 1988).

Table 2. Parameters of the partial potentials of elastic $^2$H$^3$He scattering. The boldfaced type indicates the potentials of the resonance states, which characteristics are given in the last column (Tilley et al. 2002; Ajzenberg-Selove 1988).

| $^{2S+1}L_J$ | $V_0$, MeV | $\alpha$, fm$^{-2}$ | The resonance energy $E_{cm}$, MeV | The angular momentum and parity $J^\pi$ |
|---|---|---|---|---|
| $^2S_{1/2}$, $^{2+4}D_{3/2}$ | -30.0 | 0.15 | --- | |
| **$^{2+4}D_{5/2}$** | **-31.29** | **0.09** | **3.35** | **5/2$^+$** |
| $^2P_{1/2}$ | -48.0 | 0.1 | --- | --- |
| **$^{2+4}P_{3/2}$** | **-2412.2** | **4.0** | **2.89** | **3/2$^-$** |
| **$^4S_{3/2}$** | **-34.5 (*-34.85*)** | **0.1** | **0.48** | **3/2$^+$** |
| **$^4D_{1/2}$** | **-39.88** | **0.115** | **4.14** | **1/2$^+$** |
| **$^4D_{7/2}$** | **31.99** | **0.09** | **3.06** | **7/2$^+$** |
| $^4P_{1/2}$ | -30.0 | 0.1 | --- | --- |
| **$^4P_{5/2}$** | **- 456.7** | **0.8** | **5.67** | **5/2$^-$** |

The phase shifts of the *non-resonance doublet potentials* from Table 2, given in ordinary type, are shown by the solid curves in Figure 3a. For the potential of the $^2D_{3/2}$ wave the parameters of the $^2S_{1/2}$ potential at $L = 2$ were used. In $P_{3/2}$ and $D_{5/2}$ scattering waves that are mixed along the spin channel, there are resonances that follow from the nuclear spectra shown in Figure 1 (Tilley et al. 2002; Ajzenberg-Selove 1988). Parameters of such potentials are given in bold face in Table 2 and phase shifts are shown in Figure 3a by the dashed curves.

Let us further consider the potentials for quartet scattering states whose phase shifts are shown by the solid curves in Figure 3b. If we plot the $^4S_{3/2}$ potential over the scattering phase shifts, the parameters from Table 2 can be used. The phase shift of this potential is shown in Figure 3b by the solid curve 1. Using this potential at $L = 2$ the $^4D_{3/2}$ phase shift is obtained. It is shown in Figure 3b by the solid curve 2. At 3.7 MeV in the c.m. the potential leads to a resonance with a width of 4.3 MeV in the c.m. It should be noted that in the spectra shown in Figure 1 such a resonance is not observed. Therefore, for $^4D_{3/2}$ waves we will use the $^2D_{3/2}$ scattering potential, since these states are spin-mixed with the same total momentum.

Now we recall that we use a single-channel $d + ^3$He model in which the influence of other channels is not taken into account. To substantiate the using of a single-channel model in this case, we provide interesting results obtained by Navratil and Quaglioni (2012), where they considered $d + ^3$He elastic scattering in the framework of the *ab initio* method. In Figure 3c a comparison of the theoretical calculations of the doublet and quartet $S$ scattering phase shifts performed here within the framework of the MPCM and *ab initio* calculations (Navratil & Quaglioni 2012) along with the experimental data are presented. From the figure one can see that for the $^2S_{1/2}$ phase shift all variations of the calculations for a different number of channels yield very close results, *i.e.* the influence of other channels is almost negligible. For the *ab initio* $^4S_{3/2}$ phase shift channel coupling accounting (black dashed and dotted curves) also appears very slightly.

The real agreement of the scattering phase shifts extracted from the experiment (Jenny et al. 1980), the present calculations, and *ab initio* is observed only when the channel $p + ^4$He and channel $d^* + ^3$He with the excited deuteron coupling is used by Navratil & Quaglioni (2012). In our opinion, this is a rather unexpected result, which, apparently, requires physical substantiation and interpretation. At the same time, for the purposes of our work, we can assume that the presented comparison makes it possible to draw a conclusion that the single-channel approximation employed in the present work is justified, since taking into account the two channels $p + {}^4He$ and $d^* + {}^3He$ in the *ab initio* method practically does not change the results for the scattering phase shifts in the single-channel case. It is well know that at low energies the asymptotic part of wave function plays a significant role in the most of nuclear processes, but at high energy the internal part of the wave function is very important due the strong *NN* interaction at small distances, while the latter one has a slight influence on low energy characteristics of reactions.

Let us now consider in more detail the potentials for partial waves with resonances. Figure 1 clearly shows the first resonance of $^5Li$, which is 0.482 MeV above the $^2H^3He$ channel threshold. Its excitation energy is 16.87 MeV, momentum $J^\pi = 3/2^+$, and the width is 0.267 MeV in c.m. This state can be attributed to a $^4S_{3/2}$ wave of the continuous spectrum, and its description requires a potential with FS for the diagram {5} and parameters

$V_{S3/2}$ = -34.85 MeV and $\alpha$ = 0.1 fm$^{-2}$ (5)

which lead to a resonance energy of 0.480 (1) MeV in c.m. The corresponding phase shift is shown in Figure 3b with the red dashed curve 1. For *L* = 2, the potential gives a $^4D_{3/2}$ phase shift resonance at 3.6(1) MeV in c.m. with a width of 4.1(1) MeV in c.m. Its phase shift is shown in Figure 3b with the green dashed curve 2. The potential with parameters (5) differs little from the one given in Table 2, but more accurately describes the resonance energy. At a resonance energy of 0.480 MeV, it leads to the scattering phase shift equal to 89.45°.

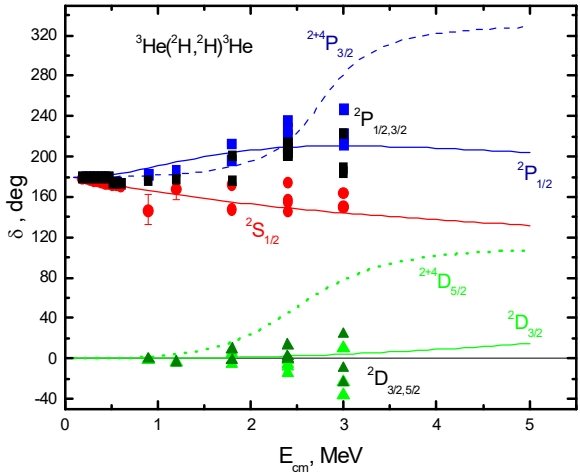
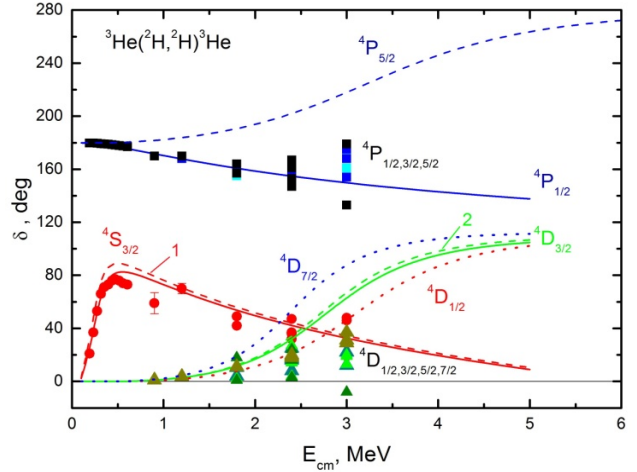

**Figure 3a.** Comparison of the $^2H+^3He$ doublet phase shifts (solid curve) mixed by orbital diagrams calculated with the potentials from Table 2 with results from (Jenny et al. 1980). Blue and black squares – results from (Jenny et al. 1980) for the $^2P_{1/2}$ and $^2P_{3/2}$ waves, green triangles – for the $^2D_{3/2}$ and $^2D_{5/2}$ waves, red dots – for the $^2S_{1/2}$ wave. The notation for several total momentums $^2P_{1/2,3/2}$ refers to the results of (Jenny et al. 1980). The preliminary results of these calculations were reported by Dubovichenko et al. (2017c).

**Figure 3b.** Comparison of the $^2H+^3He$ quartet phase shifts (solid curve), which pure by orbital diagrams calculated with the potentials from Table 2 with results from (Jenny et al. 1980). Blue, cyan and black squares – results from (Jenny et al. 1980) for the $^4P_{1/2}$, $^4P_{3/2}$ and $^4P_{5/2}$ waves, green and dark yellow triangles – for the $^4D_{1/2}$, $^4D_{3/2}$, $^4D_{5/2}$ and $^4D_{7/2}$ waves, red dots – for the $^4S_{3/2}$ wave. The notation for several full momentums $^4P_{1/2,3/2,5/2}$ is explained in Figure 3a.

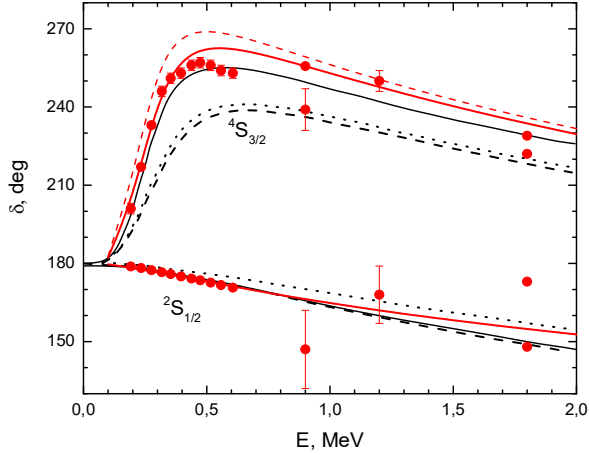

**Figure 3c.** Comparison of the calculations of the scattering *S* phase shifts for potentials in MPCM (red curves) and in *ab initio* (Navratil & Quaglioni 2012) (black curves). The dashed curve – calculation taking into account the coupling of the $p + {}^4\text{He}$ and $d + {}^3\text{He}$ channels; dotted curve – calculation **without regard** to channel coupling; solid curve – calculation taking into account the channel $p + {}^4\text{He}$ and the excited deuteron in the $d^* + {}^3\text{He}$ channel coupling.

The second resonance in the ${}^2\text{H}{}^3\text{He}$ system appears at 19.28 MeV with the angular momentum $J^\pi = 3/2^-$ and width of 0.959 MeV (Tilley et al. 2002) is laying at 19.28 MeV relative to the GS (2.892 MeV towards the channel threshold). The $P_{3/2}$ wave may match this state in doublet or quartet spin channel. The phase shift analysis presented by Jenny et al. (1980) shows no any resonance behavior of the $P_{3/2}$ waves. While, the ${}^2P$ phase shift illuminates the smooth rising in the doublet channel, its quartet partner ${}^4P$ is clearly slowly decreasing. For properly reproducing this ${}^{2+4}P_{3/2}$ resonance we need a potential with the FS and the following parameters:

$$V_{P3/2} = -2412.2 \text{ MeV and } \alpha = 4.0 \text{ fm}^{-2}. \tag{6}$$

The potential (6) provides the following resonance parameters: the position is fitted at the energy of 2.890(1) MeV in the c.m., and the corresponding width is 0.962(1) MeV in c.m. Its $P_{3/2}$ phase shift is shown in Figure 3a with a blue dashed curve, and at the resonance energy it has a value of 90.0(1)°.

In Figure 1 the first column shows the higher laying levels for $J^\pi = 7/2^+$ and $J^\pi = 5/2^+$ with the following corresponding parameters: the energy position of 19.45 and 19.71 MeV relative to the GS (or 3.06 and 3.32 MeV in c.m. relative to the channel threshold), the level width of 3.28 and 4.31 MeV in c.m. (see Table 5.3 given by Tilley et al. (2002). The strong $E1$ transitions to the GS are conditioned by direct capture from the doublet and quartet channels corresponding to the second $5/2^+$ resonance. For a correct description of the ${}^{2+4}D_{5/2}$ resonance a potential without the FS and with the following parameters

$$V_{D5/2} = -31.29 \text{ MeV and } \alpha = 0.09 \text{ fm}^{-2} \tag{7}$$

is needed.

The calculated $D_{5/2}$ scattering phase shift is shown in Figure 3a with the green dotted curve. At the resonance energy the value of this phase sift is 90.0(1)°. The calculated resonance parameters are the following: the energy position of 3.32 (1) MeV in c.m., and the width of 4.09(1) MeV in c.m. There is no resonance for the $D_{5/2}$ phases shift at such energies (Jenny et al. 1980). The ${}^4D$ phase shifts have a slight tendency to rise at 3.0 MeV in c.m. Let us mention that some our preliminary results (Dubovichenko et al. 2017c) of these calculations are different than results of the calculations shown in Figure 3a.

The level with $J^\pi = 7/2^+$ at 3.06 MeV in c.m. and a width of 3.28 MeV in c.m. can be attributed to the $^4D_{7/2}$ state. In this case the $M2$ transition to the GS is possible, the total cross sections for which will be noticeably smaller than the cross sections for the $E1$ processes. However, we give the potential without the FS for this resonance the following parameters

$V_{D7/2}$ = -31.99 MeV and $\alpha$ = 0.09 fm$^{-2}$. (8)

This potential leads to resonance at energy of 3.06(1) MeV in c.m. with a width of 3.18(1) MeV in c.m., its $D_{7/2}$ scattering phase shift is shown in Figure 3b with the blue dotted curve. At the resonance energy the phase shift has a value of 90.0(1)°.

The next level can be detected at the excitation energy of 20.53 MeV or at 4.14 MeV in c.m. with a momentum $J^\pi = 1/2^+$ and a width of 5.0 MeV in c.m., so it can be attributed to the $^4D_{1/2}$ resonance. A correct description of such resonance requires A potential with the following parameters:

$V_{D1/2}$ = -39.88 MeV and $\alpha$ = 0.115 fm$^{-2}$ (9)

This potential leads to the resonance at an energy of 4.14(1) MeV in c.m. with a width of 5.1(1) MeV in c.m., and its scattering phase shift is shown in Figure 3b by the red dotted curve. At resonance energy the phase shift has a value of 90.0(1)°.

Let us consider one more level, which lies at an excitation energy of 22.06 MeV or 5.67 MeV in c.m. with a momentum $J^\pi = 5/2^-$ and a width of 15.5 MeV in c.m. This resonance can be attributed to the $^4P_{5/2}$ state. A potential with the FS and parameters

$V_{P5/2}$ = -456.7 MeV and $\alpha$ = 0.8 fm$^{-2}$ (10)

is required to correctly describe the $^4P_{5/2}$ state.

The potential leads to resonance at the energy of 5.67(1) MeV in c.m. with a width of 15.7(1) MeV in c.m., and its scattering phase shift is shown in Figure 3b with a blue dotted curve; at the resonance energy the phase shift has a value of 90.0(1)°.

The detailed study of these levels shows that the phase shift analysis performed by Jenny et al. (1980) does not take into account the position of the resonances under consideration with large widths. The phase shift analysis is a subject to further refinement with the expansion of the energy region to 5–7 MeV in c.m. However, in order to make a detailed phase shift analysis the measurements of differential cross sections in the energy region of interest in steps of 0.3–0.5 MeV (in the region of narrow resonances the energy step should be even smaller) is required. In order to properly depict the phase shift resonance, it is required to have a step of measuring cross sections in the resonance region of not less than $\Gamma/5$. In other words, within the width of the resonance there must be at least five points of the cross sections measurement. Only in this case the resonance form of the phase shift appears quite accurately (Dubovichenko 2015b). For example, for a resonance of the potential (6) described with a width of about $\Gamma \sim 1$ MeV, a step of ~0.2 MeV is needed. For the potential (5) with resonance in the $^4S$ wave, the energy step should be of the order of 0.05–0.06 MeV.

### *3.3 Potentials for $^2H^3He$ bound states*

Now consider the potential of the GS of $^5$Li in the $^2H^3He$ channel. The $P$ states with potentials from Table 2 allowed in the doublet and quartet channels have an energy that does not agree with the binding energy of the $P_{3/2}$ and $P_{1/2}$ levels of the ground and first excited states of $^5$Li, whose spectrum is shown in Figure 1 (Tilley et al. 2002; Ajzenberg-Selove 1988). These potentials depend on two Young orbital diagrams, and the BS potential should depend only on one orbital diagram {41} (Neudatchin et al. 1992). In addition, the channel energies for the GS and FES of $^5$Li were specified above. Therefore, we carried out a refinement of the potentials of these BSs and the results are given in Table 3. In the calculations of the BS energy the exact masses of the particles were used, and the relative accuracy of the calculations for binding energy is at a level of $10^{-6}$ MeV (Dubovichenko & Dzhazairov-Kakhramanov 2015).

Table 3. New potential parameters for the $^2$H$^3$He system in pure state by Young diagrams for the GS and FES. Here the width parameter α is equal to 0.18 fm$^{-2}$. The binding energy of the FES is obtained from (Tilley et al. 2002), where 1.49 MeV above the GS is given.

| $L_J$ | $V_0$, MeV | $E_{bs}$ ($E_{exp}$ (Tilley et al. 2002)), MeV | $C_W$ | $R_{rms}$, fm |
|---|---|---|---|---|
| $P_{3/2}$ (GS) | -83.505593 | -16.388000 (-16.388) | 6.30(1) | 2.25 |
| $P_{1/2}$ (FES) | -80.485333 | -14.898000 (-14.898) | 5.74(1) | 2.26 |

In Table 3, in the third column, the experimental energy values of the levels are given in parenthesis (Tilley et al. 2002). The dimensionless asymptotic normalizing coefficients (ANC) $C_w$ are given in the last column in Table 3. They are defined according to (Plattner & Viollier 1981)

$$\chi_L(R) = \sqrt{2k_0} C_W W_{-\eta L+1/2}(2k_0 R). \tag{11}$$

In Eq. (11) $\chi_L(R)$ is the numerical GS radial WF, viz. the solution of the Schrödinger equation normalized to unit, $W_{-\eta L+1/2}(2k_0 R)$ is Whittaker function, $k_0$ is a wave number related to the channel binding energy, $\eta$ is the Coulomb parameter, $L$ is the orbital angular momentum of the bound state (Dubovichenko 2015a). The pointed ANC error is determined by its averaging over the interval from 5–6 up to 8–10 fm and shown in parenthesis. The charge root mean square radii ($R_{rms}$) for the BSs of $^5$Li in the $^2$H$^3$He channel were also calculated and are given in Table 3.

Since the ground $P_{3/2}$ state is spin-mixed, it is necessary to consider the $E$1 transitions from the doublet and quartet states of $S$ and $D$ scattering. In the framework of the model under consideration, it is impossible to explicitly isolate the $^2P_{3/2}$ and $^4P_{3/2}$ parts in the WF of the GS. So, for the calculations we will use the spin-mixed function of the $P_{3/2}$ state, which is obtained when solving the Schrödinger equation with a given GS potential from Table 3. For the scattering states quartet and mixed over the Young diagrams doublet potentials from Table 2 are used and also the resonance potentials (5)–(10).

### *3.4 Multipole Transitions*

The complete set of $E$1 and $M$1 transition amplitudes accounted in our calculations is given in Table 4. Transitions from the resonance state with the main input to the total cross sections are marked as bold. Transitions from non-resonance scattering states give a noticeably smaller contribution at low energies. The last column contains the values of the coefficient $P^2$ in expressions (2) and (3).

Table 4. Transitions accounted for calculations the total radiative $^3$He($^2$H,γ)$^5$Li capture cross section.

| № | $(^{2S+1}L_J)_i$ | The resonance energy in MeV | The angular momentum and parity $J^\pi$ | Transition | $(^{2S+1}L_J)_f$ | $P^2$ |
|---|---|---|---|---|---|---|
| 1. | $^2S_{1/2}$ | --- | | $E1$ | $^2P_{3/2}$ | 4 |
| 2. | $^4S_{3/2}$ | **16.87** | $3/2^+$ | **$E1$** | **$^4P_{3/2}$** | **4** |
| 3. | $^2P_{1/2}$ | --- | | $M1$ | $^2P_{3/2}$ | 4/3 |
| 4. | $^2P_{3/2}$ | **19.28** | $3/2^-$ | **$M1$** | **$^2P_{3/2}$** | **5/3** |
| 5. | $^4P_{1/2}$ | --- | | $M1$ | $^4P_{3/2}$ | 10/3 |
| 6. | $^4P_{3/2}$ | **19.28** | $3/2^-$ | **$M1$** | **$^4P_{3/2}$** | **22/15** |
| 7. | $^4P_{5/2}$ | **22.06** | $5/2^-$ | **$M1$** | **$^4P_{3/2}$** | **18/5** |
| 8. | $^2D_{3/2}$ | --- | | $E1$ | $^2P_{3/2}$ | 4/5 |
| 9. | $^2D_{5/2}$ | **19.71** | $5/2^+$ | **$E1$** | **$^2P_{3/2}$** | **36/5** |
| 10. | $^4D_{1/2}$ | **20.53** | $1/2^+$ | **$E1$** | **$^4P_{3/2}$** | **2/5** |
| 11. | $^4D_{3/2}$ | --- | | $E1$ | $^4P_{3/2}$ | 64/25 |
| 12. | $^4D_{5/2}$ | **19.71** | $5/2^+$ | **$E1$** | **$^4P_{3/2}$** | **126/25** |

The interaction potentials have been corroborated by the experimental data on the elastic scattering phase shifts and energy levels spectra, so, the WFs obtained as the solutions of the Schrödinger equation with those potentials account effectively the cluster system states, in particular, of the mixing by channel spin. Therefore, the total cross section of the $E1$ transition from the mixed continuous states to the spin mixed GS may be taken as simple doubling of the partial cross section as each is calculated with the same functions. However, spin algebraic factors are specified for every matrix element (2) (Dubovichenko 2015a, 2016):

$$\sigma_0(E1) = {}^2\sigma(^2D_{3/2} \to {}^2P_{3/2}) + {}^4\sigma(^4D_{3/2} \to {}^4P_{3/2}).$$

In reality, there is only one transition from the scattering state to the GS, rather than two different $E1(^2\sigma + {}^4\sigma)$ processes. The averaging procedure concerns the transitions from the $D_{5/2}$ and $D_{3/2}$ scattering states to the $P_{3/2}$ GS of $^5$Li in the $^2$H$^3$He channel. This approach, which we proposed earlier for transitions in neutron radiative capture on $^{15}$N (Dubovichenko 2016), was used for some other reactions, and allowed us to obtain reasonable results in describing the total cross sections (Dubovichenko 2016).

Thus, the total cross section of the capture process on the GS for electromagnetic $E1$ transitions is represented as the following combination of partial cross sections

$$\sigma_0(E1) = \sigma(^2S_{1/2} \to {}^2P_{3/2}) + \sigma(^4S_{3/2} \to {}^4P_{3/2}) + \sigma(^4D_{1/2} \to {}^4P_{3/2}) +$$
$$+[\sigma(^2D_{3/2} \to {}^2P_{3/2}) + \sigma(^4D_{3/2} \to {}^4P_{3/2})]/2 + [\sigma(^2D_{5/2} \to {}^2P_{3/2}) + \sigma(^4D_{5/2} \to {}^4P_{3/2})]/2$$

Since for the $M1$ transitions there is also spin mixing for some $P$ scattering waves, the total cross section is written in the same way as $E1$ transitions to the GS:

$$\sigma_0(M1) = \sigma(^4P_{5/2} \to {}^4P_{3/2}) + [\sigma(^2P_{1/2} \to {}^2P_{3/2}) + \sigma(^4P_{1/2} \to {}^4P_{3/2})]/2 +$$
$$+[\sigma(^2P_{3/2} \to {}^2P_{3/2}) + \sigma(^4P_{3/2} \to {}^4P_{3/2})]/2$$

It should be noted that $M1$ transitions from non-resonance $P$ scattering states exert a noticeable effect on the total cross sections only at energies above 300-400 keV in c.m. Thus, we have identified all the major transitions that may contribute to the total cross sections of the $^3$He$(^2$H$,\gamma)^5$Li capture process at low energies, which are treated here. Of course, the $M2$ transition is also possible, but we do not consider it because of the smallness of its cross sections.

## 4. RESULTS FOR TOTAL CROSS SECTIONS, ASTROPHYSICAL $S$-FACTOR AND REACTION RATE

In this Section, we present results obtained with the aforementioned potentials for the total cross section, the astrophysical $S$-factor, the $S$-factor screening effects, and the rate of the $^3$He$(^2$H$,\gamma)^5$Li reaction of radiation capture.

### *4.1 Total cross section*

The results of the calculated $E1$ and $M1$ radiative capture in $^2$H$^3$He cluster channel at the energies up to 5.0–6.0 MeV are presented in Figure 4. The solid red curve denotes the cross section for the $E1$ transition from the $^2S$ and $^4S$ scattering states to the GS $^{2+4}P_{3/2}$ defined by the interaction potential parameters from Table 3, while for the scattering potentials, the data from Table 2 are used. Cross sections for the $E1$ transitions from the $^2S$ wave are of few orders suppressed as it is of non-resonance behavior.

Solid violate curve in Figure 4 denotes the cross section for the $E1$ transition to the GS from the resonating $^{2+4}D_{5/2}$ waves calculated with parameters (7) and (9). This result includes all other small in value amplitudes for the non-resonating $D$ waves listed in Table 2 (see Table 4). The main contribution here is given by the $D_{5/2}$ resonance, and the contribution of the $D_{1/2}$ wave is very small. Green curve 2 in Figure 4 shows the contribution of the $M1$ transitions from the resonating $^{2+4}P_{3/2}$ and $^{2+4}P_{5/2}$ waves corresponding to potentials (6) and (10) and non-resonance set for $P$ potentials from Table 2. Figure 4 clearly shows the $P_{3/2}$ resonance at 2.89 MeV in c.m., the value of which is even larger than the resonance for the $E1$ transition, since the additional contribution is given by the $P_{5/2}$ resonance. The total cross section included all $E1$ and $M1$ transitions listed above is shown by the curve 3 in Figure 4.

In Figure 4 one can see that at energies above 400 keV the results of calculation is much lower than the available experimental data (Buss et al. 1968), and at energies less than 2 MeV they lie below the results obtained by Schroder and Mausberg (1970). This difference in the cross sections can be due to two reasons: i. in our calculations, we considered 12 different electromagnetic transitions, but perhaps some additional valuable processes were not taken into account; ii. Buss et al. (1968) and Schroder and Mausberg (1970) in the energy range from 0.4 to 2 MeV the effect of capture on the FES is not entirely excluded. In addition, the maximum of the calculated cross section in the 3 MeV region is lower than the data given by Schroder and Mausberg (1970), which can be explained by the inaccuracy of the spectra (Ajzenberg-Selove 1988) known for the authors (Schroder & Mausberg 1970), but used by us today (Tilley et al. 2002). In both cases, the results of (Buss et al. 1968) and (Schroder & Mausberg 1970) require refinement on the basis of modern methods of experimental measurements.

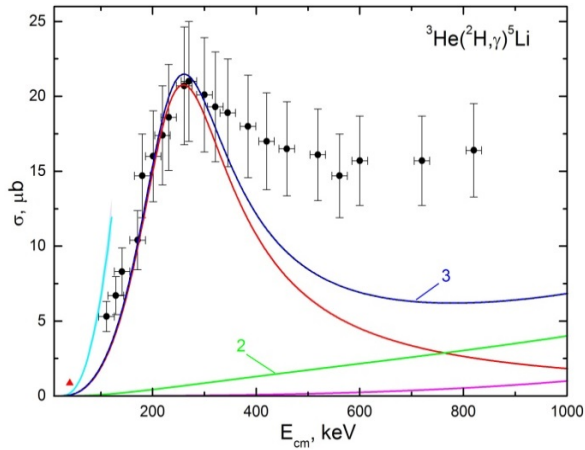 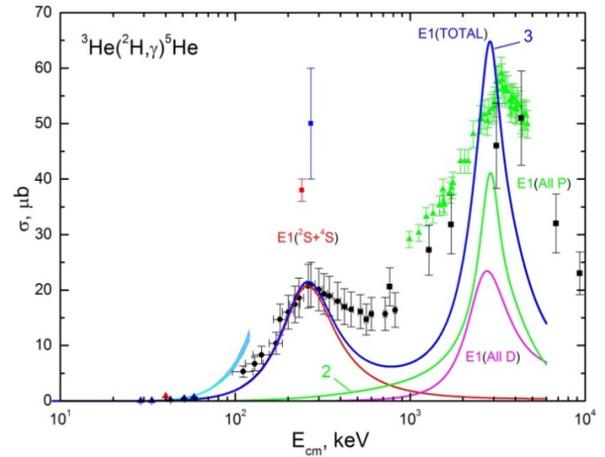

**Figure 4a.** Total cross section for $^3$He($^2$H,$\gamma$)$^5$Li bellow 1.0 MeV. Experimental data from (Buss et al. 1968) – black dotted and (Cecil, Liu & Yan 1993) – red triangle, cyan band – results from (Kiptily et al. 2006). Calculated curves correspond to the potential parameters of tables 2 and 3 and text. Two last black dots are taken from Figure 9 of the work (Buss et al. 1968).

**Figure 4b.** Total cross section for $^3$He($^2$H,$\gamma$)$^5$Li bellow 5.0 MeV. Experimental data from (Buss et al. 1968) – black dotted and (Cecil, Liu & Yan 1993) – red triangle, (Cecil, Cole & Philbin 1985; Aliotta et al. 2001) – blue triangle, (King, Meyerhof & Hirko 1972) – black square, (Schroder & Mausberg 1970) – green triangles, (Blair et al. 1954) – blue square, (Kraus et al. 1968) – red square, cyan band – results from (Kiptily et al. 2006). Two last black dots are taken from Figure 9 of the work (Buss et al. 1968).

We note that the experimental error of the cross sections given in Figure 4 was taken to be 19%, as indicated by Buss et al. (1968) for an energy of 450 keV in l.s. or 270 MeV in c.m. and the cross section 21(4) μb with an error of 15 keV for the center-of-mass energy. The data given as a function of laboratory energy by Buss et al. (1968) were recalculated by us to the c.m. energy with integer particle masses. The data given as a function of the excitation energy by Schroder and Mausberg (1970) were also recalculated to the c.m. energy with the binding energy of the $^2$H$^3$He channel of 16.4 MeV, which is given by Schroder and Mausberg (1970). The data from (Blair et al. 1954) are given for the c.m. energy of 0.27(3) MeV, at which the maximum values of the cross sections are observed in this work. The data from (Kraus et al. 1968) are given for an energy of 240 keV in the c.m. Let us note that results of calculations of the total cross section for $^3$He($^2$H,$\gamma$)$^5$Li presented in Figure 4a are obtained using a new parametrization of the potentials in contrast to our previous calculations (Dubonichenko et al. 2017c). One can find the difference of the parameters by comparing Table 3 with the results presented in Table 3 by Dubonichenko et al. (2017c).

*4.2 Astrophysical S-factor*

The astrophysical *S*-factors that characterize the behavior of the total cross section of the nuclear reaction at an energy tending to zero are determined as follows (Fowler et al. 1975,)

$$S(NJ, J_f) = \sigma(NJ, J_f) E_{cm} \exp\left(\frac{31.335 Z_1 Z_2 \sqrt{\mu}}{\sqrt{E_{cm}}}\right),$$

where σ is the total cross section of the radiative capture process in barn, $E_{cm}$ is the particle energy usually measured in keV in the center-of-mass system, μ is the reduced mass of the particles in the initial channel in amu, $Z_1$ and $Z_2$ are the charges of particles in units of elementary charge and *N* are the *E* or *M* transitions of the *J*-th multipolarity to a finite $J_f$ state of the nucleus. The value of the numerical coefficient 31.335 is obtained on the basis of the modern values of the fundamental constants (Mohr & Taylor 2005).

Now we present the results obtained for the astrophysical *S*-factor for the reaction under consideration. Figure 5 displays the total astrophysical *S*-factor (the solid curve) for the transitions to the GS due to the *E*1 and *M*1 processes in direct correspondence with the cross sections shown in Figure 4. Note, the results for the *S*-factor in Figure 5 have the same qualitative behavior as our earlier reported preliminary results (Dubovichenko et al. 2017). However, there are some variations in quantitative values. So that, *S*-factor at the energy of 6-30 keV in c.m. is still relatively stable but equal to 0.125(2) keV·b, which is still less comparing the experimental data reported by Buss et al. (1968). The error of the calculated *S*-factor shown here is obtained by averaging it over the indicated energy interval. The value of the calculated *S*-factor is 0.125 keV·b at the energy of 6 keV. At a maximum energy of 230 keV in c.m. the *S*-factor equals to 0.43 keV·b. If one uses for the $^4S_{3/2}$ scattering wave the resonance potential (5) obtained on the basis of the characteristics of the first resonance level, instead of its parameters from Table 2, it does not significantly change the magnitude and shape of the calculated *S*-factor.

We recalculated data (Buss et al. 1968) for the cross sections into *S*-factor and presented them as points in Figure 5. As we defined at minimal energies 100–200 keV its value is near 0.39 keV·b. This value in the indicated energy range can be approximated by a trivial constant energy dependence $S(E) = S_0$ with $S_0 = 0.386$ keV·b and a mean value for $\chi^2 = 0.21$. Experimental 19% errors were assumed for *S*-factor and the result is shown by the green dashed curve 1 in Figure 5.

To improve the description of the experimental data we tried the following approximating function (Canon et al. 2002)

$S(E \text{ in keV}) = S_0 + S_1 E + S_2 E^2$ (11)

but did not succeed at this very low energy region.

In what follows we implement the parametrization of the calculated *S*-factor according to the expression (11) with $S_0 = 0.12133E00$ keV·b, $S_1 = -0.12718E-04$ b, $S_2 = 0.73463E-05$ b·keV$^{-1}$ for the energy range up to 150 keV in c.m. We found the value $\chi^2$ to be 0.31 within 1% precision of the theoretical *S*-factor. The result is shown by the red dashed curve 2 in Figure 5 and is consistent with experimental data in the previously mentioned energy region.

Experimental data shown by the dotted curve in Figure 5 can be approximated by the Breit-Wigner type function

$$S(E) = a_1 + \frac{a_2}{(E - a_3)^2 + a_4^2/4} \quad (12)$$

with the following parameters $a_1 = 0.11413$, $a_2 = 7370.5$, $a_3 = 189.78$, $a_4 = 313.92$ (the energy is given in keV in c.m.). The results of this parametrization is shown by the solid curve 3 in Figure 5, $\chi^2$ is equal to 0.1 at 19% of experimental errors. At zero energy, this parametrization gives $S(0) = 0.236$ keV·b. This form of parametrization at a resonance energy of 190 keV leads to a width of 314 keV, which almost coincides with the $3/2^+$ resonance parameters (Tilley et al. 2002), given above.

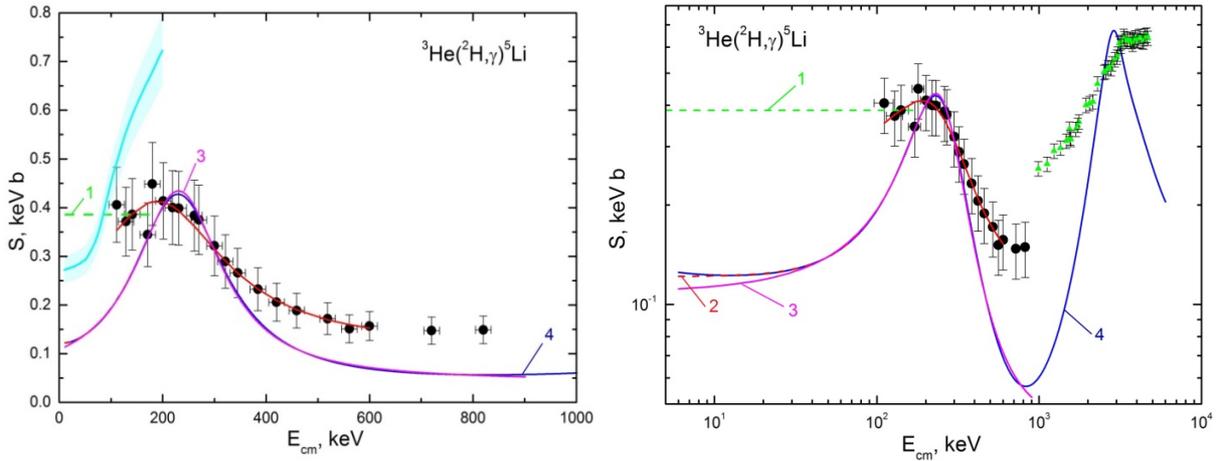

**Figure 5a.** $S$-factor data from (Buss et al. 1968) and calculations with potentials from Tables 2 and bellow 1.0 MeV. Data are the same as in Figure 4a. Cyan band – results given by Kiptily et al. (2006).

**Figure 5b.** Same as in Fig. 4a, but at the energy bellow 5.0 MeV. Data are the same as in Figure 4b.

We apply the ordinary $\chi^2$ statistics as usually was done by Dubovichenko (2015a, 2016) and defined as

$$\chi^2 = \frac{1}{N}\sum_{i=1}^{N}\left[\frac{S^a(E_i) - S^c(E_i)}{\Delta S^c(E_i)}\right]^2 = \frac{1}{N}\sum_{i=1}^{N}\chi_i^2.$$

Here $S^c$ is the original, i.e. calculated and $S^a$ is an approximate $S$-factor for the $i$-th energy, $\Delta S^c$ is the error of the original $S$-factor, which was usually taken equal to 1%, and $N$ is the number of points in the summation in the expression. As the original $S$-factor its experimental or calculated values shown in points or the solid curve 4 in Figure 5 were used, and the approximated $S$-factor is obtained on the basis of expressions (11) and (12).

We also use parametrization (12) for the calculation of the $S$-factor in the energy range up to 0.9 MeV c.m. and obtained the following parameters: $a_1 = 0.43449\text{E-}01$, $a_2 = 0.41439\text{E+}04$, $a_3 = 0.23046\text{E+}03$, $a_4 = 0.20604\text{E+}03$. Figure 5 with a violet solid curve illustrates the quality of this procedure at $\chi^2 = 7.7$ at 1% error. At 6 keV in the c.m. or zero energy the value of the approximated $S$-factor (12) is 0.11 keV·b. The parametrization leads to a resonant energy of 230 keV at a width of 206 keV, which also in a good agreement with the data reported by Tilley et al. (2002). Let us note that results of calculations for the *S*-factor presented in Figure 5a are obtained using a new parametrization of the potentials given in Table 3 in contrast to our previous calculations (Dubonichenko et al. 2017c) and in a different range of energy.

*4.3 Screening effects*

The screening effects in plasma in laboratory as well as astrophysical conditions are discussed in detail in review (Bertulani and Kajino 2016). Let us dwell only on the key points that can be applied to the reaction under consideration. The following relations for the cross sections and, accordingly, for the *S*-factors, are the generally accepted approximation for the estimation of electron screening

$$\frac{\sigma_s(E)}{\sigma_b(E)} = \frac{S_s(E)}{S_b(E)} = \frac{E}{E+U_e} \exp(\pi \eta U_e / E)$$

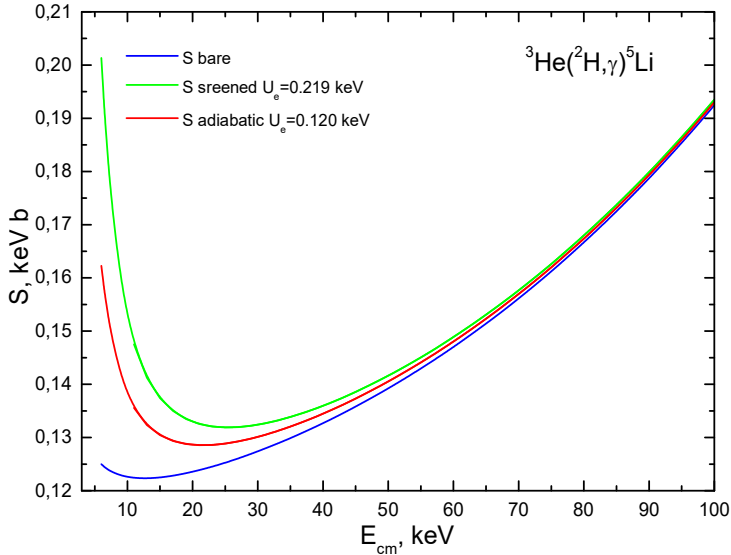

**Figure 6.** The screening effects for the astrophysical *S*-factor of the $^3\text{He}(^2\text{H},\gamma)^5\text{Li}$ capture.

Here $\sigma_s(E)$ is the electron-screened cross section, $\sigma_b(E)$ is a cross bare nuclei and $U_e$ is a constant electron screening potential energy. The Sommerfeld parameter is $\eta = \frac{Z_1 Z_2}{\hbar c}\sqrt{\frac{\mu c^2}{2E_{cm}}}$. The value of $U_e$ can be calculated analytically (*the adiabatic approximation*), or determined experimentally.

In our case, the $^3\text{He}(^2\text{H},p)^4\text{He}$ process is of interest, since it has a common initial channel with the reaction under consideration $^3\text{He}(^2\text{H},\gamma)^5\text{Li}$. For this initial channel $U_e = 120$ eV is the adiabatic limit value (Bertulani & Kajino 2016). It should be noted that for different experimental conditions $U_e$ has a different value. The direct measurements by Aliotta et al. (2001) give the values $U_e = 219\pm7$ eV for $^3\text{He}(^2\text{H},p)^4\text{He}$ and $U_e = 109\pm9$ eV for $^2\text{H}(^3\text{He},p)^4\text{He}$. Thus, for the same reaction the difference is almost two fold. In a later experimental work of Barbui et al. (2013), the screening effect turned out to be negligible. For some reason, the reference (Barbui et al. 2013) is missing in the review (Bertulani & Kajino 2016), where the significance of electron screening for astrophysical plasma is discussed in terms of *pycnonuclear ignition*.

Figure 6 illustrates the possible effect of electron screening for the process we are considering. The value of $S_b$ corresponds to the blue curve in Figure 6, e.g. $U_e = 0$, and the minimum value for $E_{cm}$ is 6 keV. It is clearly seen from Figure 6 that the screening effect strongly depends on the value of $U_e$, and it varies from 65 to 219 keV in different works (Bertulani & Kajino 2016). Obviously, until this value is determined quite accurately, it is also impossible to determine the impact of the screening effect. Therefore, it is not taken into account in calculating the reaction rate.

### *4.4 Reaction rates*

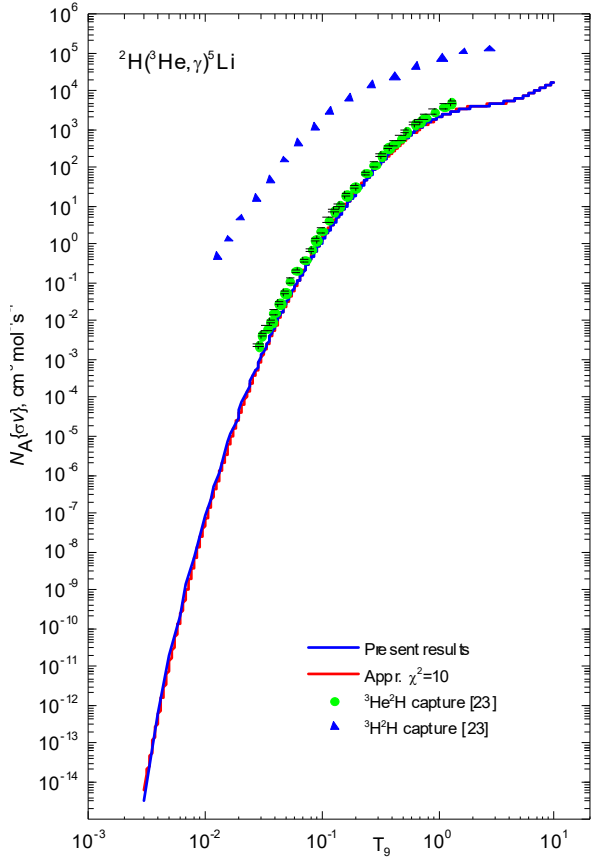

**Figure 7.** Reaction rate of the $^3\text{He}(^2\text{H},\gamma)^5\text{Li}$ radiative capture. Theoretical curves obtained with potentials from Tables 1 and 3. An explanation of the parameterization functions is given in text. Green dots are the results from (Cecil, Cole & Wilkinson 1985) for the $^3\text{He}(^2\text{H},\gamma)^5\text{Li}$ capture, the blue triangles are the results of Cecil, Cole & Wilkinson (1985) for the $^3\text{H}(^2\text{H},\gamma)^5\text{He}$ capture.

The reaction rate in $\text{cm}^3\text{mol}^{-1}\text{s}^{-1}$ units is usually defined as (Angulo et al. 1999)

$$N_A\langle\sigma v\rangle = 3.7313\cdot 10^4 \mu^{-1/2} T_9^{-3/2} \cdot$$
$$\cdot \int_0^\infty \sigma(E) E \exp(-11.605 E / T_9) dE,$$

where the energy $E$ is taken in MeV, total cross section $\sigma(E)$ in $\mu b$, the reduced mass $\mu$ in amu, and temperature $T_9$ in $10^9$ K (Angulo et al. 1999). To calculate this integral $10^4$ points of the theoretical cross section were taken in c.m. energy range from 6 keV to 10 MeV.

In Figure 7 the solid curve shows the results of our calculations for reaction rate of the $^3\text{He}(^2\text{H},\gamma)^5\text{Li}$ capture on the GS at $T_9$ from 0.003 to 10, which corresponds to the same curve in Figures 4 and 5. It should be borne in mind that the reaction rate was obtained from the calculated cross section, which slightly differs from the experimental cross section.

The dotted curve in Figure 7 display the results of Cecil, Cole & Wilkinson (1985), which have a larger value of reaction rate. This is due to the strong decrease of our calculated total cross sections in the energy range $0.5 \pm 1.0$ MeV and insufficiently accurate description of the cross sections at energies 1–6 MeV. This reaction rate is more than an order of magnitude higher than the rate of the $^3\text{He}(^2\text{H},\gamma)^5\text{Li}$ reaction considered here.

The calculated reaction rate shown on Figure 7 was approximated in the range 0.03–10.0 $T_9$ as follows (Caughlan & Fowler 1988)

$$N_A\langle\sigma v\rangle = 20986.49/T_9^{2/3} \cdot \exp(-7.13006/T_9^{1/3}) \cdot (1.0 + 126.7438 \cdot T_9^{1/3} -$$
$$- 28.83194 \cdot T_9^{2/3} + 65.13721 \cdot T_9 + 47.34928 \cdot T_9^{4/3} - 22.35306 \cdot T_9^{5/3}) - \quad (13)$$
$$- 4.94921 \cdot 10^6 / T_9^{1/2} \exp(-8.34422/T_9^{1/2})$$

The resulting curve is shown as red one in Figure 7, $\chi^2$ is equal to 10. In the approximation, the calculated points shown in Figure 7 with a blue solid curve were used. To estimate the $\chi^2$ the error was taken to be 1%.

## 5 ALTERNATIVE WAY OF $^6$LI NUCLEI FORMATION AT THE BBN

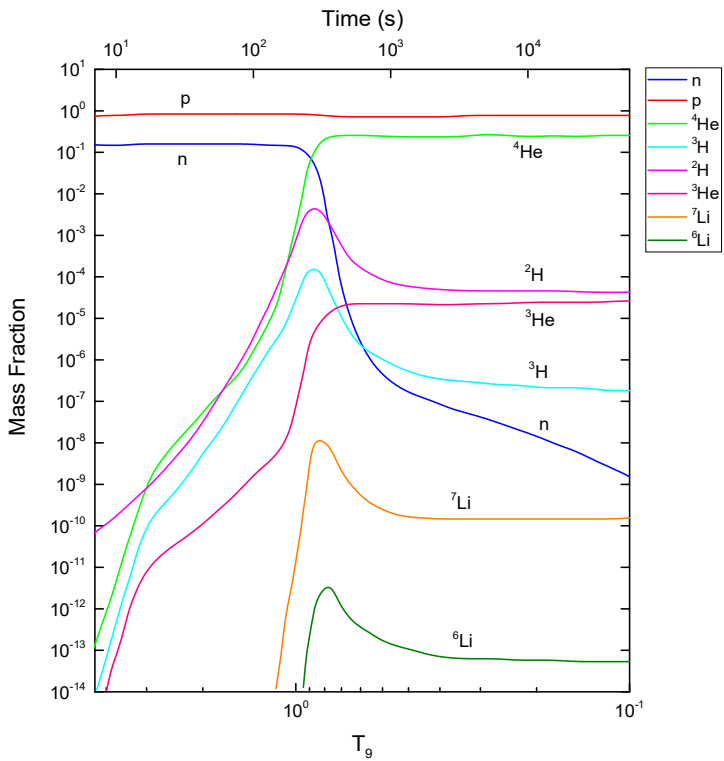

**Figure 8.** Nuclei formation mass fractions in the Big Bang thermonuclear possesses

Furthermore, we consider a possible scenario of thermonuclear astrophysical processes involving a short-lived $^5$Li isotope, which is formed as a result of the $^3$He($^2$H,$\gamma$)$^5$Li reaction. In any plasma, when a short-lived $^5$Li isotope is formed the probability of neutron capture followed by the formation of a stable $^6$Li is not zero. At first, such a chain of two reactions allows us to overcome the mass disruption at $A = 5$, leading to a stable nucleus with $A = 6$. Secondly, the reaction $^5$Li($n,\gamma$)$^6$Li leads to an additional $^6$Li formation channel, which can be considered for the explanation of the prevalence $^6$Li.

We will describe below a possible and, as it seems to us, a new scenario for the synthesis of the $^6$Li isotope in astrophysical processes and, first of all, at the Big Bang. It is required to find out whether there is a "temperature window" that would allow such an amount of $^6$Li to be accumulated during the Big Bang, which could change the balance of $^6$Li, including in the region of a sharp drop in the number of neutrons at $T_9 < 1$. The general dynamics of the synthesis of the light and lightest elements within the conditions of the standard Big Bang is represented by a graph of the participating particles fractions, shown in Figure 12 in the work of Coc et al. (2012). This graph is adapted here for our purposes and shown in Figure 8. It displays a sharp drop in the number of neutrons at $T_9 < 1$, which turn out to be bound in heavier nuclei. The number of such nuclei, for example, $^4$He and $^3$He, increases strongly at lower temperatures, since the energy of the interaction becomes so small for their breakdown.

We present below two reactions of radiative capture, which are usually considered as candidates for the role of the $^6$Li formation processes in the Big Bang and compare their rates with each other. It is believed that the radiative capture reaction

$$^4\text{He} + {}^2\text{H} \rightarrow {}^6\text{Li} + \gamma \tag{14}$$

is one of the source of $^6$Li in the Big Bang, *i.e.*, in the temperature range 50–400 keV (Bertulani & Kajino 2016). The relationship between temperature and energy is assumed to be 1.0 $T_9$ = 86.17 keV (Angulo et al. 1999).

Therefore, the energy range from 50 up to 400 keV corresponds approximately to the temperature range 0.6–4.5 $T_9$. A feature of this process is the forbiddance of the "strong" dipole $E$1 transition by the selection rules. So the reaction is due to the $E$2 transition, excepting the lowest energies, where the $E$1 process makes an appreciable contribution to the overall picture. As a result, the cross section of reaction (14) is only a few nanobarns and sharply decreases with energy decreasing due to the Coulomb barrier. The specified parametrization of this reaction rate is given by Trezzi et al. (2017)

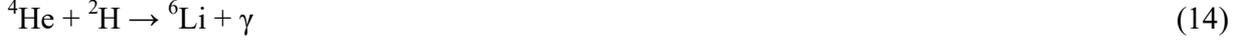

$$N_A \langle \sigma v \rangle = 20.07 T_9^{-2/3} \exp(-7.47 T_9^{-1/3})(1.0 - 4.709 T_9^{1/3} + 17.219 T_9^{2/3} - 23.839 T_9 + \\ + 19.27 T_9^{4/3} - 3.752 T_9^{5/3}) + 65.409 T_9^{-3/2} \exp(-7.565 T_9^{-1}) \tag{15}$$

This parametrization is used for comparison of the rates of radiative capture processes under consideration. There are no fundamental differences between (15) and similar parametrization from the survey (Caughlan & Fowler 1988).

Theoretical and experimental aspects of the $^6$Li formation problem in the $^4$He + $^2$H → $^6$Li + γ reaction (Lithium problem) are discussed in detail in one of the most recent publication (Grassi et al. 2017). The authors believe that despite all the efforts of accurate theoretical calculations, even when the tensor component of the $^4$He$^2$H forces (Dubovichenko 1998) is taken into account, it is possible to overcome the discrepancy with the experimentally established prevalence of $^6$Li/H ~ 1.7 10$^{-14}$ only within ~20%.

Therefore, as an additional source of $^6$Li formation in the Big Bang, the reaction (Zylstra, et al. 2016)

$$^3\text{He} + {}^3\text{H} \rightarrow {}^6\text{Li} + \gamma. \tag{16}$$

was considered. However, we could not find the parametrization of the rate of such a reaction or its theoretical calculations. In order to calculate the rate of this reaction, we perform the parametrization of its total cross sections. We use parametrization in the form of a polynomial with the Breit-Wigner term

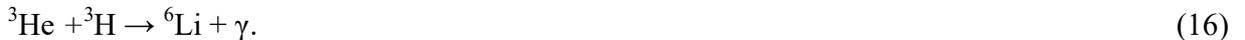

$$\sigma(E) = a_1 + \frac{a_2}{(E-a_3)^2 + a_4^2/4} + a_5 E^{b_1} + a_6 E^{b_2} + a_7 E^{b_3} + a_8 / E^{b_4} \exp(-a_9 / E^{b_5}) \tag{17}$$

All experimental data are shown in Figure 9. Basic experimental data were taken from (Young, Blatt & Seyler 1970). Here we give 23 points for the total cross sections and we took one more point from (Zylstra, Herrmann, Johnson et al. 2016) at 81 keV in c.m. The parametrization (17) with the parameters given in Table 5 leads to $\chi^2 = 0.4$, and the result of the parametrization is shown in Figure 9 with the red solid curve 1.

Table 5. The parameters of the expression (17) for parameterizing the data in Figure 9.

| $a_1$ | $a_2$ | $a_3$ | $a_4$ | $a_5$ | $a_6$ | $a_7$ | $a_8$ | $a_9$ |
|---|---|---|---|---|---|---|---|---|
| -0.66815E+02 | 0.92167E+03 | 0.18011E+01 | 0.64950E+01 | -0.26955E+01 | -0.31659E+00 | 0.16969E+00 | 0.90199E+04 | 0.60666E+01 |

| $b_1$ | $b_2$ | $b_3$ | $b_4$ | $b_5$ |
|---|---|---|---|---|
| 0.63402E+00 | 0.65888E+00 | 0.22340E+01 | 0.15001E+01 | 0.62889E+00 |

All parameters given in Table 5 are varied independently to obtain a minimum value of $\chi^2$. From Table 5 we see that the power of $b_4$ is 3/2, as is usually assumed in similar calculations (Caughlan & Fowler 1988), the powers of $b_1$, $b_2$ and $b_5$ are close to 2/3, and the power of $b_3$ is approximately equal 2. But if we take powers, excepting the one of $b_4$, to be equal to these integers, this leads to an increase in $\chi^2$ of up to 1.4. Moreover, if we perform additional variation of the remaining parameters $a_i$, then, despite the small value of $\chi^2 = 0.3$, the approximated cross section becomes negative even at energies below 50 keV, which strongly affects the shape of the reaction rate. Therefore, in order to calculate the reaction rate, we used the parametrization (17) with the parameters given in Table 5. This parametrization leads to negative cross sections at energies below 5 keV, therefore the calculation of the cross section is limited by this energy value. In Figure 9, the green solid curve 2 shows the cross section obtained from the parameterization (17) and calculated in the energy range from 0.005 to 10 MeV, which is directly used later for calculations of the rate of reaction (16).

Furthermore, another reaction can be considered. The following process can be estimated as an additional $^6$Li synthesis channel

$$n + {}^5\text{Li} \to {}^6\text{Li} + \gamma. \tag{18}$$

Obviously, a direct measurement of the cross section of this process is impossible. However, according to detailed balance, this cross section can be estimated from data for the photodisintegration reaction $^6$Li($\gamma,n$)$^5$Li with binding energy $E_{51} = 5.67$ MeV. The arguments in favor of the fact that the cross section of the reaction (18) can be significant is the absence of a Coulomb barrier, and also a centrifugal barrier, since a "strong" dipole $E1$ transition is realized from scattering $S$-waves.

For further analytical calculations of the reaction (18) rate two variants of the cross section parameterization were obtained. In Figure 10a the solid curve shows the following version of the parameterization

$$\sigma(E) = a_1 + \frac{a_2}{(E - a_3)^2 + a_4^2/4} + a_5 E \tag{19}$$

with the parameters $a_1 = -0.62090$E-01, $a_2 = 0.10927$E+03, $a_3 = 0.64065$E+01, $a_4 = 0.79845$E+01,

$a_5$= 0.27650E+00 with $\chi^2$ = 0.023.

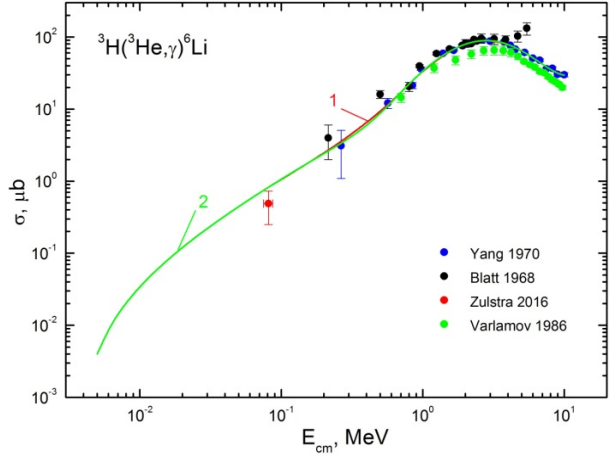

**Figure 9.** Radiative $^3$H($^3$He,$\gamma$)$^6$Li capture cross sections. The experimental data are as follows: (Blatt et al. 1968) – •, (Young, Blatt & Seyler 1970) – •, (Zylstra, Herrmann, Johnson et al. 2016) – •. For comparison, the results of (Varlamov et al. 1986) are given (•). The experimental data are taken from the mentioned above works and the database (Nuclear Reaction Database (EXFOR), 2013). Solid curves show present different parameterizations (17).

These parameters were obtained as a parametrization of the data of (Varlamov et al. 1986) taken from Ref. (Nuclear Reaction Database (EXFOR), 2013) and shown with the dots in Figure 10. However, this parameterization does not properly describe the behavior of the cross section at thermal energy. The cross section reaches a plateau with a value of 1.9 microbars for any, arbitrarily low energy. Since the thermal cross section for this reaction was not measured by anyone, we can assume for it the value of 40 mb, as for $^7$Li (Firestone & Revay 2016). The use of this data on thermal neutrons for $^7$Li is completely justified, because in the ground state this nucleus has the same quantum numbers $J^\pi = 3/2^-$ as $^5$Li, while it is reasonable, for the first estimate, to assume that they have similar rms sizes.

For a correct description of the thermal cross sections, it is required to change the form of the parametrization (19) as

$$\sigma(E) = a_1 + \frac{a_2}{(E-a_3)^2 + a_4^2/4} + a_5 E + a_6/\sqrt{E} + a_7/E \qquad (20)$$

with parameters $a_1$ = -0.62090E-01; $a_2$ = 0.10823E+03; $a_3$ = 0.63957E+01; $a_4$ = 0.79517E+01; $a_5$ = 0.27950E+00; $a_6$ = -0.14339E-01; $a_7$ = 0.10142E-02 and $\chi^2$ = 0.0216. The green curve in Figure 10b is obtained from the parametrization (20), taking into account the reference to the thermal neutron capture cross section. Hence one can see that the parametrized cross section well conveys the results reported by Varlamov et al. (1986) and describes the data at the thermal energy. However, one must bear in mind that at medium energies below the data given by Varlamov et al. (1986), there is a "plateau", which is not characteristic of such cross sections (Dubovichenko 2016).

Furthermore, in Figure 11 one can see the reaction rates of some radiative capture processes which can lead to synthesis of $^6$Li or can be an intermediate stage for its synthesis. The blue solid curve shows the results for the capture reaction rate of $^3$H($^3$He,$\gamma$)$^6$Li obtained on the basis of our the parameterization (17). Let us once again pay attention to the fact that such a parametrization leads to negative cross sections at energies below 5 keV. Therefore, the cross sections are cut off at this energy, which apparently leads to a more rapid decrease in the reaction rates at low temperatures. The black solid curve shows the rate of the neutron capture on $^5$Li for the case of the parametrization (19), and the light blue curve – for the parameterization (20), which makes it possible to describe the cross sections at thermal energies.

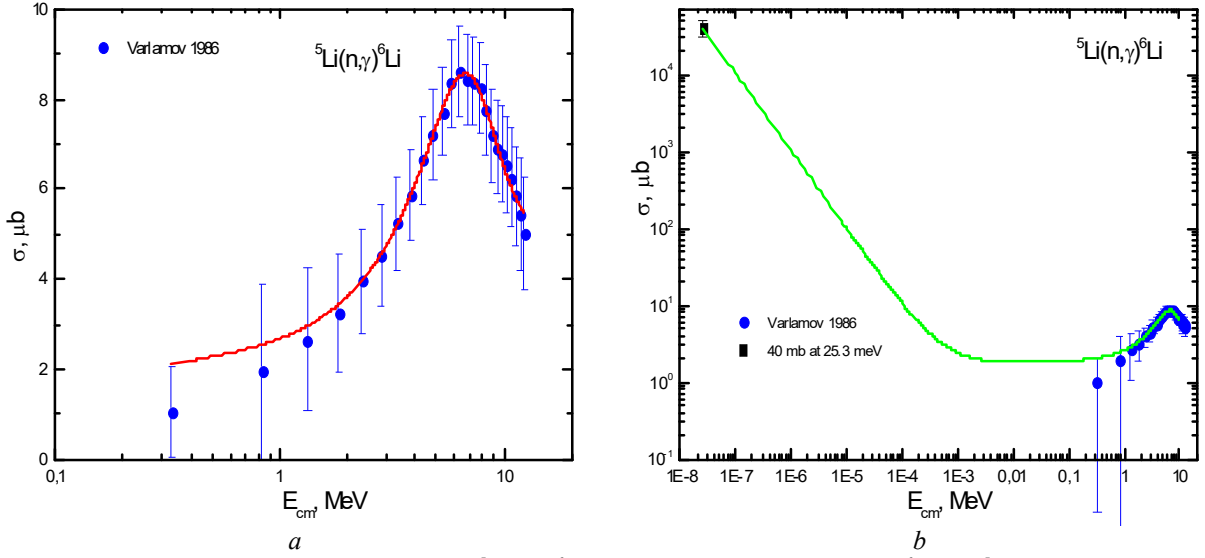

**Figure 10.** The cross section for the radiative $^5$Li$(n,\gamma)^6$Li capture based on the data for $^6$Li$(\gamma,n)^5$Li photodisintegration (Varlamov et al. 1986). The experimental data of (Varlamov et al. 1986). Curves – parametrization of the present work: *a* – without reference to thermal cross sections $\sigma_{th}$; *b* – with reference to $\sigma_{th}$ for $^7$Li.

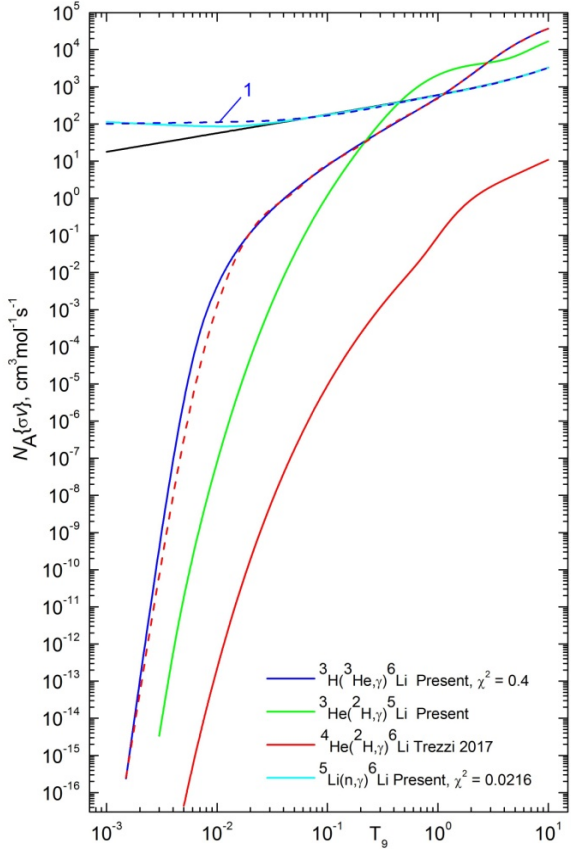

**Figure 11.** Different radiative capture reactions rates

It is easy to see an increase in the reaction rate at low temperatures if we take into account the thermal cross sections for the $^5$Li$(n,\gamma)^6$Li capture. The red curve is the rate of the capture $^4$He$(^2$H$,\gamma)^6$Li reaction, obtained from the parametrization (15) (Trezzi et al. 2017). The green curve shows our calculations of the rate of the $^3$He$(^2$H$,\gamma)^5$Li capture reaction, which can be parametrized with the expression (13) from the previous section.

As follows from Figure 11, the rate of the "priority" reaction $^4$He + $^2$H → $^6$Li + $\gamma$ is less than all the others. Therefore, the question about the role of the $^3$He + $^3$H → $^6$Li + $\gamma$ process, as a possible candidate for amplifying the $^6$Li yield arose. The answer is contained in the comparison of the reaction rates presented in Figure 11 and particles fractions in the process of BBN in Figure 8. Against the background of a large fraction of alpha particles, the deuterium fraction decreases with decreasing temperature. The synthesis of $^3$He is accelerated and to some extent compared with the amount of $^2$H, starting from $T_9$~1. In the same region, a maximum is observed for tritium $^3$H, the fraction of which, however, sharply decreases with time. As a result, the conclusions of Zylstra et al. (2016) on the minor contribution of this reaction in comparison with (14) are understandable.

It follows from Figure 11 that the reaction rate of the $^5$Li synthesis (green curve) with the participation of deuterium and $^3$He at $T_9$~1 is comparable and even greater than the process (16), and the deuterium fraction exceeds the tritium fraction by one order. In Figure 11 it is clearly shown that the rate of reaction $^5$Li(n,γ)$^6$Li is greater than for $^3$H($^3$He,γ)$^6$Li and $^3$He($^2$H,γ)$^5$Li, and even greater than for $^4$He($^2$H,γ)$^6$Li reactions. Therefore, it can be assumed that in the $T_9$~1 region, which corresponds to a large neutron concentration, we can expect the real contribution of the reaction $n + {}^5$Li → $^6$Li + γ to the $^6$Li synthesis.

Now we give the parametrizations of $^5$Li($n$,γ)$^6$Li and $^3$H($^3$He,γ)$^6$Li capture reactions rates. In the first case, it was possible to find only a polynomial form of the parametrization

$$N_A \langle \sigma v \rangle = \sum_{i=1}^{N} a_i E^{i-1} \quad (21)$$

with the parameters from Table 6. This form at $N = 7$ leads to $\chi^2 = 3.0$. Here and below, for calculations of $\chi^2$, the error was assumed to be 1%. The results of the approximation are shown in Figure 11 with the dashed curve.

Table 6. The parameters for expression (21)

| $a_1$ | $a_2$ | $a_3$ | $a_4$ | $a_5$ | $a_6$ | $a_7$ |
|---|---|---|---|---|---|---|
| 100.2674 | 717.7027 | -296.7566 | 93.84265 | -15.13806 | 1.2058 | -0.0374 |

To parameterize the rate of the second reaction, another form

$$N_A \langle \sigma v \rangle = a_1 / T_9^{a_2} \exp(-a_3 / T_9^{a_4})(1.0 + a_5 T_9^{a_6} + a_7 T_9^{a_8} + a_9 T_9^{a_{10}} + a_{11} T_9^{a_{12}}) + \\ + a_{13} / T_9^{a_{14}} \exp(-a_{15} / T_9^{a_{16}}) \quad (22)$$

was used which leads to $\chi^2 = 32.4$. The results of the approximation are shown in Figure 11 with the dashed curve 1 with parameters given in Table 7. As seen from Figure 11, the rate of the $^3$H($^3$He,γ)$^6$Li capture reaction has an unusual form. Perhaps this is a consequence of the overestimated cross section obtained from the parametrization (17) in the low-energy region. Therefore, in expression (22), it is required to vary not only the coefficients of the $T_9$ powers, but also the values of the powers themselves.

Table 7. The parameters for expression (22)

| $a_1$ | $a_2$ | $a_3$ | $a_4$ | $a_5$ | $a_6$ | $a_7$ | $a_8$ |
|---|---|---|---|---|---|---|---|
| 74.07273 | 1.8375 | 4.95476 | 0.4065 | 2183.953 | -1.69893 | -3714.616 | -1.56543 |
| $a_9$ | $a_{10}$ | $a_{11}$ | $a_{12}$ | $a_{13}$ | $a_{14}$ | $a_{15}$ | $a_{16}$ |
| 907.838 | 3.00098 | -324.5072 | 3.31375 | 11793.31 | 2.28774 | 2.48388 | 0.53456 |

Based on the obtained results, a more consistent evaluation of the role of the $n + {}^5\text{Li} \rightarrow {}^6\text{Li} + \gamma$ process in the BBN and in stars is desirable in the future. For this purpose, all necessary cross section parameterizations and reaction rates are obtained in this paper. In addition, our simple estimates based on the mass fraction distribution shows qualitatively that the role of the $n + {}^5\text{Li} \rightarrow {}^6\text{Li} + \gamma$ reaction requires a further study for astrophysical processes in conditions of high neutron concentration, which have a different temperature regime compared to the BBN.

We present calculations for the total cross sections, astrophysical $S$-factor, and reaction rates for ${}^3\text{He}({}^2\text{H},\gamma){}^5\text{Li}$ radiative capture in the framework of the potential cluster model with forbidden states using a single channel approach. We show that the the integral characteristics of ${}^3\text{He}({}^2\text{H},\gamma){}^5\text{Li}$ radiative capture can be reproduced with very high accuracy. Our results are promising and pave the way for further microscopic analyses of this process. However, it is well known that the ${}^3\text{He}({}^2\text{H},p){}^4\text{He}$ and ${}^3\text{H}({}^2\text{H},n){}^4\text{He})$ reactions are strongly dominant. We are well aware of the classical works within the method of resonating groups on the investigation of the $\alpha + N$ scattering channels taking into account the coupled-channels for five-nucleon systems. This formalism found wide acceptance following its application to calculations of various polarization characteristics (Wildermuth and Tang, 1977). Indeed, there are indications of very subtle effects associated with channels coupling for polarization characteristics. Today the Gaussian expansion method to accurately solve the Schrödinger equations, which includes heavy calculations for bound, resonant and scattering states of three- to five-body systems (see review: Hiyama and Kamimura 2018 and references) is used for coupled-channels calculations. Our calculations can be further improved by including additional $\alpha + N$ channels by performing coupled-channels calculations. The description of the five-nucleon system using the potential model within the Schrödinger equations with modern nucleon-nucleon potentials is a challenging issue and beyond the scope of the present research.

One should pay attention to the fact that the variation in the data on integral sections, which we use for astrophysical calculations, is quite large. In Section 2 we did our best to reasonably and maximally coordinate both the total cross sections and yields of the processes. We relied on this compilation of the data that served as a criterion for the reliability of our calculations. One can see that the experimental data can be reproduced within a single-channel approach. In our opinion the coupled-channels calculations will not dramatically change the resulting pattern.

In this paper we make the assumption related to the existence of the two step mechanism ${}^3\text{He}({}^2\text{H},\gamma){}^5\text{Li} \rightarrow {}^5\text{Li}(n,\gamma){}^6\text{Li}$ in formation of ${}^6\text{Li}$ as one of the options to address the lithium abundance within the BBN model. The role of short-lived isotopes in astrophysical thermonuclear processes is quickly becoming a popular subject of experimental and theoretical research, so in this pioneering work, we tried to identify some problematic aspects of these calculations regarding the lack of information in the literature for the considered processes. In addition, a role of two-step processes, which are extremely difficult to study in laboratory conditions, but which, nevertheless, occur in natural plasma, requires clarification. These processes either make a certain contribution to the scenario of stellar plasma evolution as a whole, or their role should be considered insignificant. This question is open, and we have demonstrated one of ways to address its solution.

# 6. CONCLUSION

In conclusion, we note that the main goal of our studies is to determine the role of the $^3$He($^2$H,$\gamma$)$^5$Li radiative capture reaction in the balance of processes involving deuterons that occur under natural and laboratory conditions in plasma. This reaction was not considered in the review (Caughlan & Fowler 1988), which presents the parameterization of reaction rates involving light and lightest nuclei. The contribution of this work to this important compilation consists in calculating the rate of the $^3$He($^2$H,$\gamma$)$^5$Li process based on a certain nuclear model and its corresponding parametrization (13).

### *Nuclear physics aspects*

Comparatively simple-channel model representations (namely, the use of MPCM in this work) succeeded to obtain the theoretical results in general agreement with the available experimental data for the *S*-factor or total cross section of the radiative $^3$He($^2$H,$\gamma$)$^5$Li capture. The larger value of minimum in the cross sections at energies $0.4 \pm 2$ MeV, in comparison with the experiment, can be due to the fact that we do not take into account any additional electromagnetic transitions, in spite of we took into account 12 such processes. However, there may be another explanation. In the experimental data for capture on the GS a contribution from the transition to FES is present. All the main available experimental data, reviewed in Sec. 2, were obtained and published about 45 to 60 years ago and obviously require clarification.

### *Nuclear astrophysics aspects*

From the point of view of further application of the results obtained here in astrophysical problems, we can indicate the following:

1. Simple parametrizations of the considered radiative $^5$Li($n,\gamma$)$^6$Li and $^3$H($^3$He,$\gamma$)$^6$Li capture reactions cross sections and their rates are obtained.
2. The rates of these two processes and the rate of the $^3$He($^2$H,$\gamma$)$^5$Li capture reaction considered here are compared.
3. The possible contribution of the neutron capture on $^5$Li to the formation of a stable $^6$Li is considered.
4. It has been shown qualitatively that the neutron capture on $^5$Li formed at $^3$He($^2$H,$\gamma$)$^5$Li capture in the temperature range of the order of $1.0T_9$ at the BBN, can make a significant contribution to the processes of primary accumulation of a stable $^6$Li.

On the basis of all results obtained here it is clear that now it is required to make quantitative calculations of the contribution of such reactions to the accumulation of the $^6$Li nucleus at BBN in stars and other astrophysical processes.

# ACKNOWLEDGMENTS

This work was supported by the Ministry of Education and Science of the Republic of Kazakhstan (Grant No. BR05236322) titled "Study reactions of thermonuclear processes in extragalactic and galactic objects and their sybsystems" through the Fesenkov Astrophysical Institute of the National Center for Space Research and Technology of the Ministry of Defence and Aerospace Industry of the Republic of Kazakhstan (RK).